\documentclass{aa}
 \usepackage{txfonts} 
 \usepackage{natbib} 
 \usepackage{graphicx} 
\usepackage{aalongtable}

 \bibpunct{(}{)}{;}{a}{}{,} 

\begin{document}

\title{IMAGES IV\thanks{Intermediate MAss Galaxy Evolution Sequence, ESO programs 174.B-0328(B), 174.B-0328(F), 174.B-0328(K), 66.A-0599(A)}: Strong evolution of the oxygen abundance in gaseous phases of intermediate mass galaxies since $\mathrm{z}\sim0.8$Ó }
\titlerunning{IMAGES IV: Strong evolution of the oxygen abundance}
\authorrunning{M. Rodrigues et al.}

\author{M. Rodrigues \inst{1,2}
 \and F. Hammer \inst{1} 
 \and H.Flores \inst{1} 
 \and M. Puech \inst{3,1}
 \and Y.C. Liang \inst{4}
 \and I.  Fuentes-Carrera \inst{1} 
 \and N. Nesvadba \inst{1}   
 \and M. Lehnert  \inst{1}      
  \and Y. Yang  \inst{1}
  \and P. Amram \inst{5}               
 \and C. Balkowski \inst{1}
 \and C. Cesarsky \inst{1}   
 \and  H.Dannerbauer \inst{6} 
 \and  R. Delgado \inst{1,7}
 \and  B. Guiderdoni \inst{8}  
 \and A. Kembhavi \inst{9} 
 \and   B. Neichel \inst{1} 
 \and   G. {\"O}stlin \inst{10}     
 \and L. Pozzetti \inst{11}    
 \and C.D. Ravikumar \inst{12}   
  \and  A. Rawat \inst{1,9} 
  \and  S. di~Serego~Alighieri \inst{13} 
  \and  D. Vergani \inst{14}  
 \and J. Vernet  \inst{3} 
  \and  H. Wozniak \inst{8} 
  }

\institute{
GEPI , Observatoire de Paris, CNRS, University Paris Diderot ; 5 Place Jules Janssen,  92195 Meudon, France 
\and CENTRA, Instituto Superior Tecnico, Av. Rovisco Pais 1049-001 Lisboa , Portugal 
\and ESO, Karl-Schwarzschild-Strasse 2, D-85748 Garching bei M\"unchen, Germany
\and National Astronomical Observatories, Chinese Academy of Sciences, 20A Datun Road, Chaoyang District, Beijing 100012, PR China
\and Laboratoire d'Astrophysique de Marseille, Observatoire Astronomique de Marseille-Provence, 2 Place Le Verrier, 13248 Marseille, France
\and MPIA, K{\"o}nigstuhl 17, D-69117 Heidelberg, Germany
\and IFARHU-SENACYT, Technological University of Panama, 0819-07289 Panama, Rep. of Panama
\and Centre de Recherche Astronomique de Lyon, 9 Avenue Charles Andr\'e, 69561 Saint-Genis-Laval Cedex, France
\and Inter-University Centre for Astronomy and Astrophysics, Post Bag 4, Ganeshkhind, Pune 411007, India
\and Stockholm Observatory, AlbaNova University Center, Stockholms Center for Physics, Astronomy and Biotechnology, Roslagstullsbacken 21, 10691 Stockholm, Sweden
 \and INAF - Osservatorio Astronomico di Bologna, via Ranzani 1, 40127 Bologna, Italy
\and Department of Physics, University of Calicut, Kerala 673635, India
\and INAF, Osservatorio Astrofisico di Arcetri, Largo Enrico Fermi 5, I-50125, Florence, Italy
 \and IASF-INAF - via Bassini 15, I-20133, Milano, Italy
}

\date{Received  /
Accepted }
\abstract {Intermediate mass galaxies ($>$$10^{10}~M_\odot$) at z$\sim$0.6 are the likeliest progenitors of the present-day, numerous population of spirals. There is growing evidence that they have evolved rapidly since the last 6 to 8 Gyr ago, and likely have formed a significant fraction of their stellar mass, often showing perturbed morphologies and kinematics.}  {We have gathered a representative sample of 88 such galaxies and have provided robust estimates of their gas phase metallicity.}
{For doing so, we have used moderate spectral resolution spectroscopy at VLT/FORS2 with unprecedented high S/N allowing to remove biases coming from interstellar absorption lines and extinction to establish robust values of $R_{23}=([OII]\lambda3727 + [OIII]\lambda \lambda 4959,5007)/H\beta$.}{We definitively confirm that the predominant population of z$\sim$ 0.6 starbursts and luminous IR galaxies (LIRGs) are on average, two times less metal rich than the local galaxies at a given stellar mass. We do find that the metal abundance of the gaseous phase of galaxies is evolving linearly with time, from z=1 to z=0 and after comparing with other studies, from z=3 to z=0. Combining our results with the reported evolution of the Tully Fisher relation, we do find that such an evolution requires that $\sim$$30\%$ of the stellar mass of local galaxies have been formed through an external supply of gas, thus excluding the close box model. Distant starbursts \& LIRGs have properties (metal abundance, star formation efficiency \& morphologies) similar to those of local LIRGs. Their underlying physics is likely dominated by gas infall probably through merging or interactions.}{Our study further supports the rapid evolution of z$\sim$0.4 --1 galaxies. Gas exchanges between galaxies is likely the main cause of this evolution.}

\offprints{ myriam.rodrigues@obspm.fr} 

\maketitle
\section{Introduction}
There is a growing consensus that most of the decline of star-formation density since z=1 is related to the strong evolution in the intermediate-mass galaxy population, with $M_{\mathrm{stel}}$ from 1.5 to 15$\times10^{10}\mathrm{M_\odot}$ \citep{2005A&A...430..115H, 2005ApJ...625...23B}. At intermediate redshift these galaxies are progenitors of present day spirals. This work belongs to a series of studies on intermediate-mass galaxies at z$\sim$0.6, in the frame of the large program IMAGES (Intermediate MAss Galaxy Evolution Sequence). In Paper I \citep{2008A&A...477..789Y}, we present the 2D kinematics of a representative sample of 65 galaxies and put into evidence the high fraction of non-relaxed galaxies at this redshift. In Paper II \citep{2008A&A...484..159N}, we establish a morpho-kinematical classification and find a strong evolution of the fraction of well-relaxed spiral rotating disk along the last 6 Gyrs. We also point out that intermediate-mass galaxies have double their stellar mass since z$\sim$0.6 according to the evolution of the Tully-Fisher relation found in Paper 3 \citep{2008A&A...484..173P}. As a whole, this series of papers have shown that intermediate-mass galaxies have performed a strong evolution of their kinematical and morphological properties since z$\sim$0.6. These results reveal the agitated history of the spiral disk galaxies over the last 6 Gyrs. 

The evolution of the metal content of the gas in galaxies is an useful tool to probe various galaxy evolution scenarios. In fact, the stellar mass-metallicity relation (M-Z) can help to disentangle the contribution of several processes taking place during galaxy evolution such as star-formation history, outflow powered by supernovae or stellar winds and infall of gas by merger or secular accretion. In the local Universe the M-Z relation has been widely studied since \citet{1979A&A....80..155L}. \citet{2004ApJ...613..898T} performed a reliable estimation of oxygen abundance and stellar mass for 53\,000 star-forming galaxies from the SDSS. Several studies have characterized the relation at higher redshift: $0 < z <1$ \citep{2004ApJ...617..240K, 2006A&A...447..113L, 2005ApJ...635..260S, 2007IAUS..235..408L}, 1$\leq$z $\leq$2 \citep{2008ApJ...678..758L, 2006ApJ...644..813E,2004ApJ...612..108S, 2006ApJ...639..858M} and z$\geq$3 \citep{2008arXiv0806.2410M}. A significant evolution is found as a function of redshift: z$\sim$0.7 galaxies show on average a gaseous abundance with a factor of two less than local counterparts at a given stellar mass \citep{2006A&A...447..113L}. Such a trend is also found at higher redshift. However, the metal abundance of the gas in galaxies is estimated for all these samples from strong line calibration as $R_{23}$ which uses [\ion{O}{iii}], [\ion{O}{ii}] and $H\beta$ or $N2$, which uses [\ion{N}{ii}] and $H\alpha$. Several parameters affect the measurement of line ratios: extinction and underlying Balmer absorption in the case of $R_{23}$, the unknown contribution of primary nitrogen production as a function of the metallicity for $N2$, and for both parameters the spectral resolution. \citet{2004A&A...417..905L} emphasized the necessity of a spectral resolution over $R>1\,000$ and a S/N$>$10 to recover robust physical quantities from spectra. There are still few studies reaching the spectral quality necessary to estimate reliable metal abundance for intermediate redshift galaxies. The data from the Keck Redshift Survey are limited by the absence of flux calibration, making impossible to estimate the extinction. Moreover, metallicity estimated using the equivalent widths produce systematically higher metal abundance \citep{2006A&A...447..113L}. \citet{2005ApJ...635..260S} observed 56 galaxies from the Gemini Deep Deep Survey with flux-calibrated spectra, but S/N of the data were insufficient to measure extinction for individual galaxies. 

The aim of this paper is to obtain a robust M-Z relation for a reasonably large sample of intermediate mass galaxies at z$\sim$0.6. In this paper, we have followed the methodology proposed by \citet{2006A&A...447..113L} consisting of a precise estimate of the extinction and underlying Balmer absorption in order to obtain reliable $R_{23}$ abundance determination. The high quality spectra were taken with the VLT/FORS2. The paper is organized as follows: Sect. 2 describes the observational data and discusses the completeness of the sample. In Sect. 3, we outline the methodology followed to estimate extinction, star-formation rates and metallicities, as well as determinate the contribution of AGN. Results concerning the evolution of the M-Z relation and the effect of galaxy morphology are given in Sect. 4. In Sect. 5, we discuss systematic effects and the implication of our results on the close box scenario. We summarize our conclusions in Sect. 6. 

Throughout this paper we adopt a $\Lambda$-CDM cosmological model ($\mathrm{H_0}$=70$\mathrm{km~s^{-1}Mpc^{-1}}$, $\Omega_\mathrm{M}$=0.3 and $\Omega_\mathrm{\Lambda}$=0.7). All magnitudes used in this paper are in the AB system, unless explicitly noted otherwise. The adopted solar abundance is $12+log(O/H)$=8.66 from \citet{2004A&A...417..751A}.

\section{IMAGES observation}
 
\subsection{VLT/FORS2 Observations}
Galaxies were observed with the VLT/FORS2 in the Chandra Deep Field South (CDFS) in the context of the ESO large program "IMAGES". We randomly selected 270 galaxies in 4 fields of 6\arcmin8\,$\times$\,6\arcmin8. Observations were performed in MXU mode, using two holographic grisms, GRIS 600RI+19 and GRIS 600z+23, at spectral resolution of 6.8\,$\AA$ and covering a maxima wavelength range from 5\,120\,$\AA$ to 10\,700\,$\AA$  . The slit masks were prepared using the GOODS catalogue \citep{2004ApJ...600L..93G} and FORS2 pre-images. The arrangement of the slits has been optimized to improve the number of objects observed. 270 slits have been placed among which 241 objects have $\mathrm{I_{AB}}<$23.5\,mag. The minimum exposure time of each object in each grism is around 3 hours. \\
Spectra have been extracted and wavelength calibrated using the IRAF\footnote{IRAF is distributed by the National Optical Astronomical Observatories, which is operated by the Association of Universities for Research in Astronomy,Inc., under cooperative agreement with the National Science Foundation.} package. The spectra were flux calibrated using photometric standard stars. We have compared the spectrophotometry to HST/ACS photometry in V, I, Z bands and V, I, R in EIS catalogue and found a good agreement. 

\subsection{Samples}
\begin{figure} \resizebox{\hsize}{!}{\includegraphics{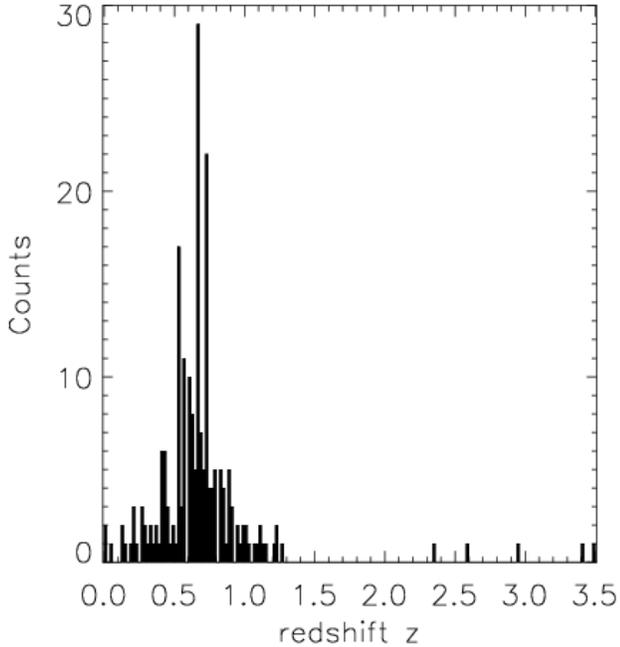}} \caption{Redshift distribution (bin=0.02) of the FORS2--IMAGES sample, including the classes 2(secure), 1(insecure) and 9(single line).} \label{hist_z} \end{figure}

From the 241 target objects with $\mathrm{I_{AB}}<$23.5\,mag from the FORS2--IMAGES sample, we have identified 231 targets. For 10 targets we have not been able to perform an identification because of instrumental problems (target falling outside the slit or bad sky extraction). Each redshift has been associated with a quality flag denoting the reliability of the identification, see description in \citet{2007A&A...465.1099R}. We have classified objects into 4 types: galaxy, split into emission line galaxies (ELG) and absorption line galaxies (ALG); quasar and stars. Table 1 shows the number and type of objects in the different redshift quality classes.
\begin{table} 
\label{table_obs} 
\centering
\begin{tabular}{c c c c c c} 
\hline
Class & Galaxy& Star & QSO & Total & \%\\ 
\hline
2&169&19&4&192&83\\
1&32&-&-&32&14\\
9&7&-&-&7&3\\
\hline 
Total& 208&19&4&231&-\\
\hline 
\end{tabular} 
\caption{ 
Number and type of objects in the redshift quality classes: secure (2) for spectra contain more than two strong features, insecure (1) when only one features was detected but with not very reliable supporting features, and single emission line (9) when the spectra had a single emission line without any features }
\end{table}
From the 270 slits of the FORS2--IMAGES, we have estimated the redshift for 240 objects, giving a success rate of $\sim$88\%. For objects with $\mathrm{I_{AB}}<$23.5\,mag the success rate reached $\sim$96\%.
In Fig.\ref{hist_z}, we have plotted the redshift histogram for all galaxies and QSO. The distribution have a range of [0.01-- 3.499] and a median value of 0.667. Two main peaks are visible in the redshift distribution at z=0.670, z=0.735. These peaks correspond to structures in the CDFS field, already identified in \citet{2007A&A...465.1099R}.\\
 
 \begin{table} 
\label{table_sample} 
\centering
\begin{tabular}{c | c c c c c c | c} 
\hline 
  & E1 & E2 & E3 & E4 & E5 & Abs &Total \\ 
\hline
IMAGES&4&65&20&31&49&43&212\\
Sample A&4&65&5&-&-&-&74\\
\hline 
\end{tabular} 
\caption{ Number of galaxies (208 galaxies + 4 QSO) by spectral type in the IMAGES and selected sample: [\ion{O}{ii}], $\mathrm{H\beta}$, [\ion{O}{iii}] and $\mathrm{H_\alpha}$ emission lines detected (E1); [\ion{O}{ii}], $\mathrm{H\beta}$, [\ion{O}{iii}]  detected (E2); [\ion{O}{ii}] out of the wavelengh range (E3); $\mathrm{H\beta}$ or/and [\ion{O}{iii}] out of range (E4); one or two emission lines detected (E5); galaxies without emission lines (Abs). The E3 galaxies correspond to low-redshift objects. In this work we have only selected the 5 galaxies with z$>$0.4}
\end{table}
From the  $\mathrm{I_{AB}}<$23.5\,mag objects we have selected 74 galaxies having [\ion{O}{ii}], $\mathrm{H\beta}$, [\ion{O}{iii}] emission lines, in order to evaluate their metallicities. Table 2 describe the construction of the working sample by spectral lines availability. 
From the sample of 74 galaxies, we have excluded 6 objects because of bad spectrophotometry or wavelength calibration problems and 10 objects with $\mathrm{H\beta}$ or both [\ion{O}{iii}] lines corrupted by strong sky emission or absorption lines. Finally, our sample, hereafter sample A, is composed by 58 galaxies with a median redshift of 0.7. The S/N of $\mathrm{H\beta}$ line has a mean value of 40. 

The CDFS field has the advantage of being a widely studied field. The Great Observatories Origins Deep Survey (GOODS) \citep{2004ApJ...600L..93G} provides deep multiwavelength observations. HST/ACS and EIS photometries \citep{2001A&A...379..740A} have been used to check the spectrophotometry and to calculate the aperture correction. The IR luminosity of galaxies has been estimated from the mid-IR catalogue of \citet{2005ApJ...632..169L} and using the procedure of \citet{2001ApJ...556..562C}. Observations in X-ray from the \textit{Chandra X-Ray Observatory 1 Ms} observations \citep{2002ApJS..139..369G} and radio detection from the Australia Telescope Compact Array 1.4 GHz \citep{2006AJ....131.1216A} have been used to identify Active Galactic Nucleus (AGN), see Sect 3.3. The rest frame magnitude in J,K-band and the luminosity at 2800$\AA$ have been derived by modeling galaxy SEDs using ISAAC and ACS multi-band photometry. \citet{2008A&A...484..173P} have evaluated that the random uncertainty of this method is less than 0.2$\sim$mag in $M_K$. A catalogue of Sample~A with RA, DEC, z, $\mathrm{I_{AB}}$ from the ACS, $M_K$, $M_J$, $L_{IR}$ is given in Table \ref{table1}. 

We have completed our sample with 30 galaxies from \citet{2006A&A...447..113L}, referred hereafter as sample B. This $I_{AB}<$22.5\,mag galaxies have been randomly selected in Canada France Redshift Survey (CFRS) 3h, Ultra Deep Survey Rosat (UDSR) and Ultra Deep Survey FIRBACK (UDSF) and observed with VLT/FORS2/MXU.
This sample has $M_\mathrm{K}$ measurement and mid-IR detection with ISOCAM, see \citet{2005A&A...430..115H} and \citet{2004A&A...423..867L} for a complete description. 

\subsection{Completeness}
\begin{figure} \resizebox{\hsize}{!}{\includegraphics{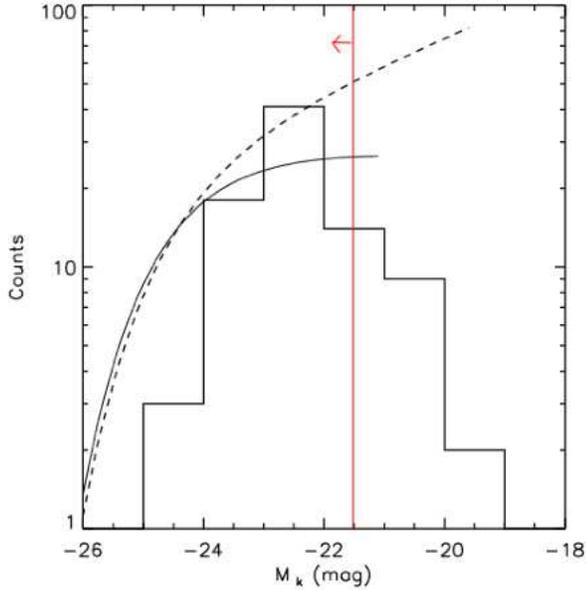}} \caption{Number counts (in logarithmic scale) of selected galaxies versus AB absolute magnitude in K-band. The black histogram refers to sample~A+B galaxies. The luminosity function derived from \citet{2003A&A...402..837P} at $\mathrm{z} = 0.50$ is plotted in dashed line and at $\mathrm{z}=1$ in solid line. The red vertical line represents the limit of 85\%-completeness.} \label{mk05} \end{figure}

In this work, we have combined samples A and B and called the whole sample: sample~A+B with a total of 88 galaxies. The two samples of galaxies span the same redshift range. In terms of absolute magnitude, sample B is composed of galaxies brighter than sample A. However, the combination of this two samples gives a sample representative of the intermediate mass galaxies at z$\sim$0.6. Fig. \ref{mk05} shows the distribution of the K-band absolute magnitudes of sample A+B and the objects with redshift identification in the FORS2--IMAGES sample verifying $\mathrm{I_{AB}}<$23.5\,mag. We have compared, on the same plot, the luminosity distribution to the luminosity function at redshift of 0.5 and 1 from \citet{2003A&A...402..837P}. Kolmogorov-Smirnov tKS tests support that sample A+B follows the LF at a 85\% probability in the redshift range z=0.4 and z=0.98, and with MK$<-$21.5. The corresponding stellar mass range of completeness is $\log(M_{stellar}/\mathrm{M_{\odot}})>10$. The aim of the IMAGES large program is to investigate intermediate mass galaxies with $M_{stellar}> 1.5\times10^{10}\,\mathrm{M_\odot}$, which represents 76\% of our sample. Then, the incompleteness of the sample under $\mathrm{M_{K}}=-21.5$ does not affect our conclusions.

\section{Methodology and data analysis}
The following section describes the methodology of data analysis for the sample~A. The description for the sample~B is given in \citet{2004A&A...423..867L} and \citet{2006A&A...447..113L}. The methodology used for sample~A is very similar to the one used for sample~B, thus the two sample are homogeneous. 

\subsection{Flux measurement}
\begin{figure*} \centering \includegraphics[width=17cm]{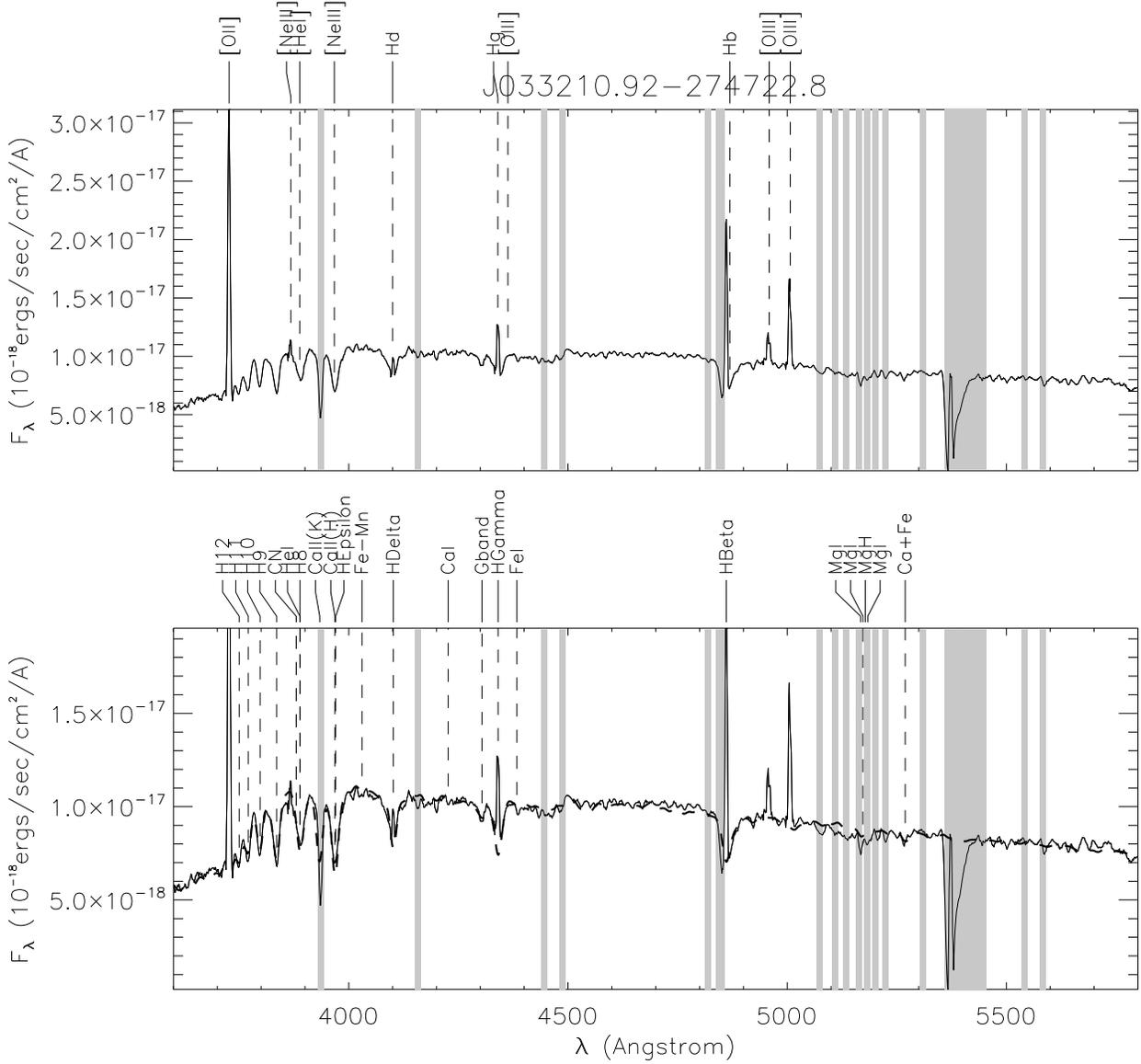} \caption{Rest-frame spectrum of one of the sample~A galaxies, J033210.92-274722.8. Grey boxes delimit the wavelength region where spectra may be affected by strong sky lines. \textbf{Top panel}: Spectrum with the location of strong emission lines. \textbf{Bottom panel}: Spectrum and synthetic spectra using \citet{1984ApJS...56..257J} templates plotted in dashed line. The position of the absorption lines are marked in dashed vertical lines. } \label{fit_cont} \end{figure*}

In order to derive metal abundance in galaxies with sufficient accuracy, it is necessary to estimate the extinction and the underlying Balmer absorption. 
We have corrected the Balmer emission lines from underlying stellar population. For each galaxy, we have fitted the observed spectrum, including its continuum and absorption lines, with a linear combination of stellar libraries. We have used a set of 15 stellar spectra from the \citet{1984ApJS...56..257J} stellar library, including B to M stars (e.g B, A, F, G, K and M) with stellar metallicity. Spectra and stellar templates have been degraded to the same spectral resolution of 8\,$\AA$. The smoothing of the spectra emphasises the absorption features used in the fit. Only the continuum and absorption lines have been smoothed in the galaxy spectra. We have convolved the continuum, except at the location of emission lines, using the software developped by our group \citep{2001ApJ...550..570H}.  The best fit has been done with the STARLIGHT software \citep{2005MNRAS.358..363C}. A \citet{1989ApJ...345..245C} reddening law has been assumed. A rest-frame spectrum of one typical galaxy of our sample, J033210.92-274722.8, is given in Fig \ref{fit_cont}.  

We have chosen to use stellar templates rather of the usually used SSP models of  \citet{2003MNRAS.344.1000B}, hereafter BC03. Indeed, BC03 models overestimate the absorption in the $\mathrm{H\beta}$ line. \citet{2007MNRAS.381..263A,2005MNRAS.358..363C} have suggested that the origin of this biais is due to the STELIB library used in the BC03 models. The STELIB spectra have an excess of flux in both side of the $\mathrm{H\beta}$  when compare to model spectrum. As the synthetic spectrum is not use to retrieve physical properties of the stellar population, we have adopted a simple stellar library in order to better control the parameters of the fit. 

After subtracting the stellar component, we have measured the flux of emission lines using the SPLOT package. When the [\ion{O}{iii}] $\lambda 4959$ emission line was not detected, the flux was assumed to be 0.33 times the [\ion{O}{iii}]$\lambda 5007$ (the ratio of the transition probability). 

\subsection{ A reliable estimate of the extinction }
We have measured extinction by two methods in the sample~A: Balmer decrement and $\mathrm{IR/H\beta}$ energy balance \citep{2004A&A...423..867L}. Comparing the two values, we can check the reliability of our methodology and avoid systematic effects on abundances due to bias in extinction evaluation. 

For 37 objects of sample~A, the $\mathrm{H\gamma}$ line has been detected and we could derive extinction in the gas phase directly from the spectra, using the observed $\mathrm{H\gamma/H\beta}$ ratio. Assuming a case B recombination with a density of $100\,\mathrm{cm^{-3}}$ and an electronic temperature of 10\,000\,K, the predicted ratio is 0.466 for $I(\mathrm{H\gamma})/I(\mathrm{H\beta})$ \citep{1989agna.book.....O}. Using the following relation: 
\begin{center}
\begin{equation}
\left(\frac{I_{\mathrm{H\gamma}}}{I_{\mathrm{H\beta}}}\right)_{\mathrm{obs}}=\left(\frac{I_\mathrm{{H\gamma0}}}{I_{\mathrm{H_\beta0}}}\right)_{\mathrm{intr}} 10^{-c(f(\mathrm{H\gamma})-f(\mathrm{H\beta}))}\, ,
\end{equation}
\end{center}
we have determined the dust extinction $c$. The extinction parameter $A_V$ is calculated following \citet{1979MNRAS.187P..73S}: $A_V=E(B-V)R=(c\,R)/1.47$ where $R=3.1$. The median extinction of the sample is $A_V=1.53$. 

Because of large uncertainties related to the measurement of the $\mathrm{H\gamma}$ line, we needed to verify the quality of our derived extinction. Another method to evaluate $A_V$ is from the infrared and optical SFR. Indeed, when dust extinction is taken into account, the $SFR_{\mathrm{H_\alpha}}$ is in good agreement with $SFR_{\mathrm{IR}}$ \citep{2004A&A...415..885F}. The ratio of the uncorrected extinction $SFR_{\mathrm{H\alpha}}$ and $SFR_{\mathrm{IR}}$ allows to estimate the amount of extinction. The two SFR have been computed assuming the \citet{1998ARA&A..36..189K} calibrations and both use the same Salpeter's initial mass function (IMF) \citep{1955ApJ...121..161S}. The $L_{\mathrm{IR}}$ is estimated from the \citet{2001ApJ...556..562C} relation and the $24\,\mu m$ observations \citep{2005ApJ...632..169L}. The $L_{\mathrm{H\alpha}}$ has been estimated from the flux of $\mathrm{H\beta}$ and assuming $\mathrm{H_\alpha}/\mathrm{H\beta}$=2.87 \citep{1989agna.book.....O}. The $\mathrm{H\beta}$ flux have been corrected by an aperture factor derived from photometric magnitudes at $I_{\mathrm{AB}}$ and $V_{\mathrm{AB}}$ and spectral magnitudes. Finally, the ratio has been corrected using the average interstellar extinction law:
\begin{equation}
A_{V}(\mathrm{IR})=\frac{3.1}{1.66}\log{\frac{SFR_{\mathrm{IR}}}{SFR_{\mathrm{H_\alpha nc}}}} \, ,
\end{equation}
The median extinction in the sample for $\mathrm{IR/H\beta}$ estimation is $A_V=1.71$. In the case of the objects without IR detection, we have used the detection limit of 0.08\,mJy of the MIPS catalogue to find an upper limit for $L_{\mathrm{IR}}$ and so an upper extinction limit $A_V(\mathrm{IR_{lim}})$. 

\begin{figure} \resizebox{\hsize}{!}{\includegraphics{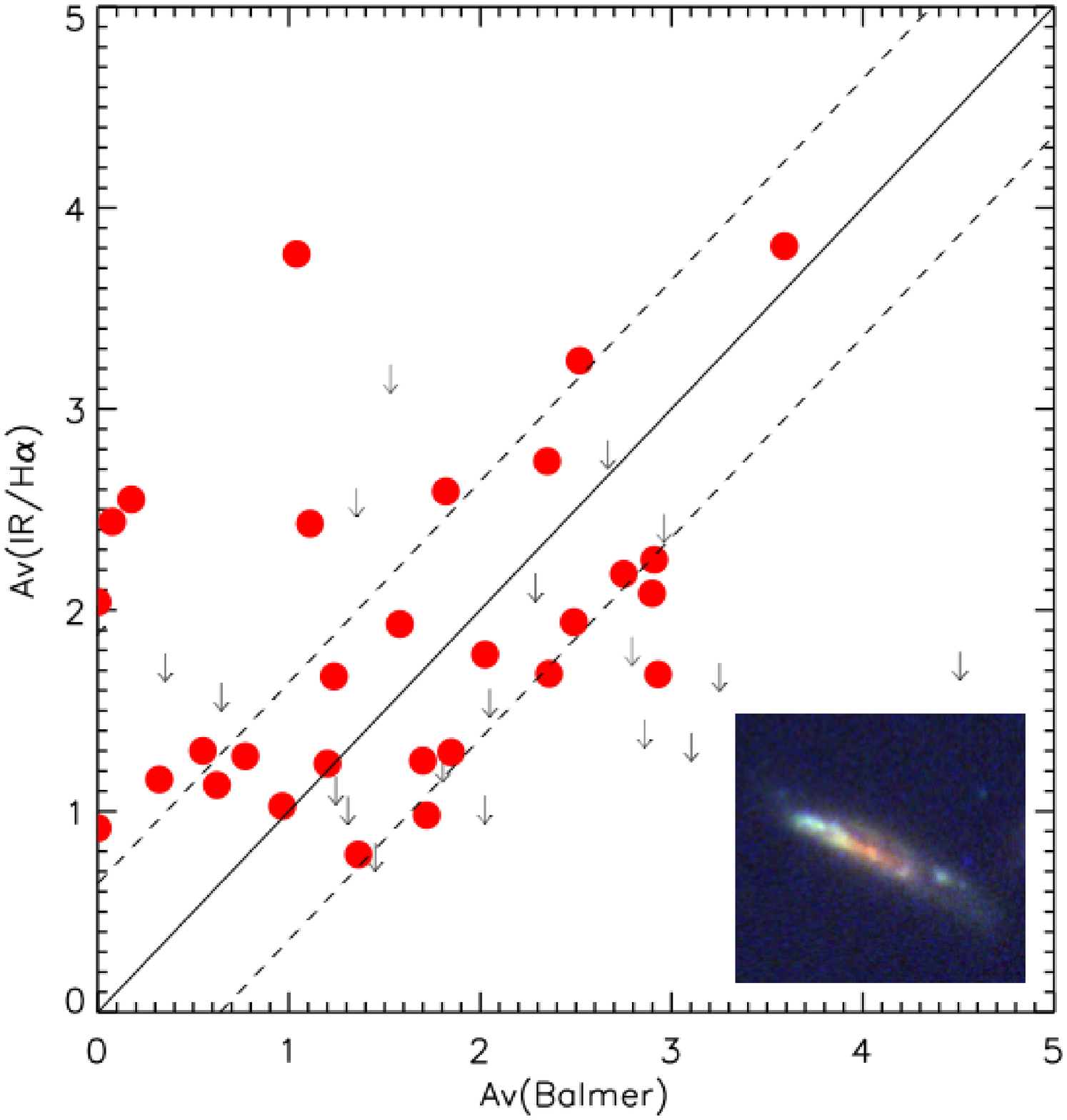}} \caption{Relation between the extinction $A_V$ values derived by Balmer decrement and by $SFR_{\mathrm{IR}}$ and $SFR_{\mathrm{H_\alpha nc}}$ ratio. We plot the 37 galaxies from Sample~A having reliable $\mathrm{H\gamma}$ detection and the 13 galaxies from Sample~B with $A_V$(Balmer) and $A_V$(IR) measurements. The objects with infrared detection are plotted in filled red circles. The arrows are the 21 objects in the Sample~A with only $A_V$(IR) upper limit estimation. The two dashed lines refer to the results with $\pm$0.64\,rms. The down left image in the graph is the combining ACS/HST image in B,V and I band of the edge-on galaxy J033212.39-274353.6. } \label{av} \end{figure}

In fig \ref{av} we compare the extinction given by the two methods. We plot in the same figure the objects from sample~A+B with $A_V$(Balmer) and $A_V$(IR) measurements. The same methodology has been used to determinate $A_V$(Balmer) and $A_V$(IR) in Sample~B, see \citet{2004A&A...423..867L} for a complete description of the method. Most galaxies fall within the $\pm$0.64~rms 1-$\sigma$ dispersion. This result is consistent with \citet{2004A&A...415..885F} and \citet{2004A&A...423..867L}. We notice that the median extinction of the sample~B is higher than sample~A: $A_V$(Balmer)=1.82 and $A_V$(IR)=2.18. This difference between the two samples can be due to the selection of the sample~B: galaxies are more luminous than in sample~A and detected in near IR. 

Some discrepancies between the two measurements of extinction are due to geometrical properties, as J033212.39-274353.6, J033232.13-275105.5 and CFRS 03.0932 which show $A_V$(IR) greater than $A_V$(Balmer). Indeed, these 3 objects are edge-on galaxies. In such case, a large fraction of the disk is hidden by the dust in the disk plane. The detected optical Balmer lines just trace the star formation of a few optical-thin HII regions lying on the periphery of the galaxy in the line of sight of the observer. The consequence is an underestimated extinction value. The infrared radiation is less affected by dust so it comes from deeper region of the galaxy. Thus, we expect to have higher extinction values when estimated with the IR flux. 
 
For non-edge-on galaxies, we have assumed that the difference between the two measurements is statistical and not intrinsic: i.e. the two extinctions are statistically equal. Hereafter, we assume $A_V$ as the mean value between $A_V$(IR) and $A_V$(Balmer) for the objects with IR detection and the mean between $A_V(\mathrm{IR_{lim}})$ and $A_V$(Balmer) for galaxies with an IR upper limit. For objects without $\mathrm{H\gamma}$ detection we have used $A_V$(IR) or $A_V(\mathrm{IR_{lim}})$ when $L_{\mathrm{IR}}$ is not available.  For edge-on galaxies, the difference between the two extinction estimates is physical as explained above. Since our aim is to correct the optical spectra, we have used $A_V$(Balmer) for these galaxies. The comparison between the two extinction estimates allows to minimize systematic errors. In fact, $A_V$(IR) tends to give higher extinction values because the mid-IR light is less affected by dust. On the other hand, the Balmer decrement may underestimate the extinction because it follows only massive stars population and regions without dust-obscured star formation. Adopting the mean values between the two methods allows to reduce the systematic errors of each one. \\

\subsection{Searching for possible AGN contamination}
Before studying metallicities in galaxies we have to identify the objects whose lines are affected by contamination from an AGN. This identification is essential because the AGN processes affect the [\ion{O}{iii}] emission lines and so any metallicity estimate based in this line.  
First, we have eliminated from Sample~A two galaxies harboring AGN spectral features, like broad \ion{Mg}{ii} and Balmer lines: J033208.66-274734.4 and J033230.22-274504.6. These two galaxies are very compact and are both X-ray detected. Broad Balmer lines suggest that these galaxies are Seyfert~1 galaxies.

\begin{figure} \resizebox{\hsize}{!}{\includegraphics{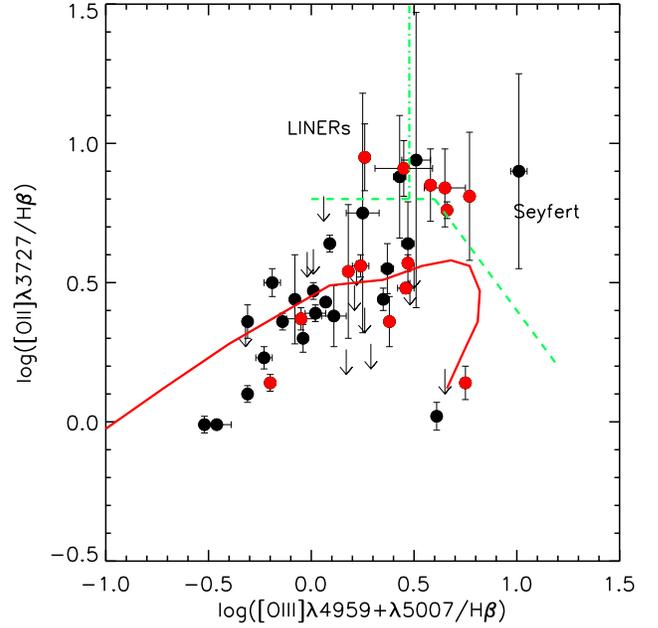}} \caption{Diagnostic diagram for the Sample~A. Galaxies with only $A_V(\mathrm{IR_{lim}})$ are plotted with down arrows. The objects with log([\ion{Ne}{iii}]$\lambda$3868/[\ion{O}{ii}]$\lambda$3727) $>$ -1.3  are plotted in red dots. The solid line shows the theoretical sequence from \citet{1985ApJS...57....1M} for extra-galactic HII regions. The dashed line shows the photo-ionization limit for a stellar temperature of 60\,000\,K and empirically delimits the Seyfert~2 and LINER from the \ion{H}{II} regions. The dot-dashed line shows the demarcation between Seyfert~2 and LINERs from \citet{1989agna.book.....O}. } \label{diagnostic} \end{figure}

Then, the diagnostic diagram log([\ion{O}{ii}]$\,\lambda$3727/$\mathrm{H_{\beta}}$) vs log([\ion{O}{iii}] $\lambda\lambda$4959, 5007/$\mathrm{H_{\beta}}$) was used to distinguish the \ion{H}{II} region-like objects from LINERs and Seyfert galaxies, see Fig. \ref{diagnostic}. We have found 10 objects in the LINER and Seyfert~2 region of the excitation diagram: J033214.48-274320.1, J033219.32-274514.0, J033222.13-274344.5, J033223.06-274226.3, J033224.60-274428.1, J033229.32-275155.4, J033236.72-274406.4, J033240.04-274418.6, J033243.96-274503.5, J033245.51-275031.0. From these objects we discard 5 galaxies which fall out of the \ion{H}{II} region, even if we account for error bars: J033219.32-274514.0, J033222.13-274344.5, J033240.04-274418.6, J033243.96-274503.5, J033245.51-275031.0.  The other 5 galaxies are very close to the limit of star-forming galaxies to classify them definitely as AGN. We have chosen to keep them and we have used a different symbol for these objects in the figures. 

We search for evidence of shock processes in our galaxy sample by the presence of the emission line $[\ion{Ne}{iii}]\lambda3868$. Five galaxies present log([\ion{Ne}{iii}]$\lambda$3868/[\ion{O}{ii}]$\lambda$3727) $>$ -1.3 and fall in the Seyfert~2 area. One of them,  J033236.72-274406.4 have IR and X-ray detection. 

\begin{figure} \resizebox{\hsize}{!}{\includegraphics{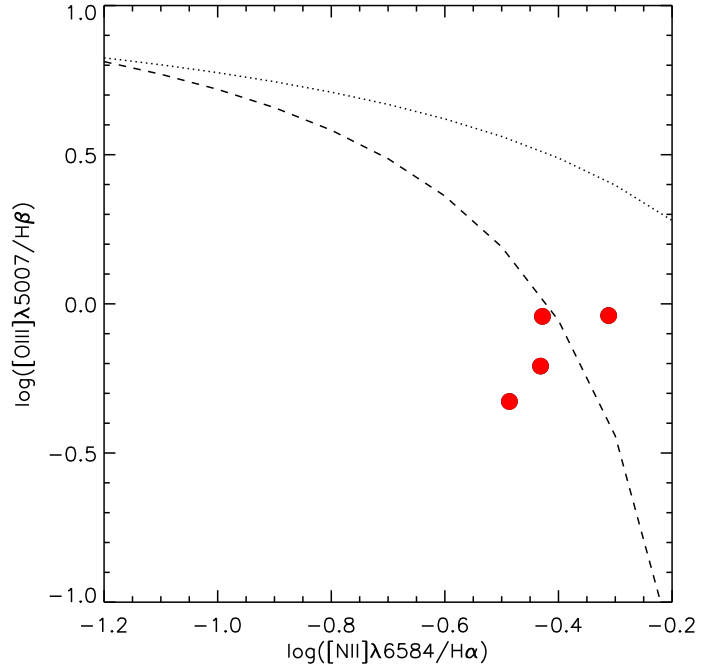}} \caption{Diagnostic diagram for the Sample~A. The dots line shows the  limit of the region occupied by star-forming galaxies from \cite{2001ApJS..132...37K}. The dashed line represents the empirical demarcation separating star-fromation galaxies from AGN from \citet{2003MNRAS.346.1055K}. } \label{diagnosticNII} \end{figure}

For 4 objects, $\mathrm{H_{\alpha}}$ and [\ion{N}{ii}] measurements are available and we have checked the log([\ion{N}{ii}]$\lambda$ 6584/$\mathrm{H_{\alpha}})$ vs log([\ion{O}{iii}] $\lambda\lambda$4959, 5007/$\mathrm{H_{\beta}}$) diagnostic diagram, see fig. \ref{diagnosticNII}: all of them fall in the \ion{H}{II} region delimited by \citet{2002ApJS..142...35K}. 


\subsection{Estimating the metallicity of the ionized gas}
We have evaluated the metal abundance in the warm ionized gas. Such estimate reflects the current metallicity of the gas from which the next generation of stars will form. Accurate metallicity measurements in the gas phase require the determination of the electron temperature $T_{\mathrm{e}}$, usually given by the ratio of auroral to nebular line flux, such as [\ion{O}{iii}]$\lambda\lambda$4959,5007/[\ion{O}{iii}]$\lambda$4363. Unfortunately, in metal-rich environments the [\ion{O}{iii}]$\lambda$4363 line is too weak to be detected. Furthermore, even at low metallicities this auroral line is hardly detectable in low S/N spectra from high-z galaxies. 

To overcome this difficulty, several estimators based on strong emission lines are available in literature. We have used the $R_{23}$ parameter defined by \citet{1979MNRAS.189...95P} as $R_{23}$=([\ion{O}{ii}]$\lambda$3727 + [\ion{O}{iii}]$\lambda\lambda$4959,5007)/$\mathrm{H\beta}$. There are a large number of theoretical and empirical calibrations linking $R_{23}$ to oxygen abundance \citep{2001A&A...369..594P,2002ApJS..142...35K,2004ApJ...613..898T, 2005ApJ...631..231P, 2007A&A...473..411L}. Abundances determined by different indicators give substantial biases and discrepancies. For example, the difference between calibrations based on electronic temperature and on photo-ionization model can reach 0.5\,dex \citep{2007A&A...473..411L,2008ApJ...674..172R,2008arXiv0801.1849K}. 
As the aim of this work is to compare metallicities at different redshifts, we have adopted the calibration from \citet{2004ApJ...613..898T} to avoid bias related to calibration:
\begin{equation}
12+log(O/H)= 9.185-0.313x-0.264x^2-0.321x^3 \, , 
\end{equation}
where $x=\log{R_{23}}$. Indeed, we want to compare sample~A+B with the SDSS sample from which the Tremonti calibration is based. 

The $R_{23}$ calibration presents some limitation in abundance determination, namely it is double valued with $12+log(O/H)$. At low metallicity $R_{23}$ scales with metal abundance. But from $12+log(O/H)=8.3$, gas cooling occurs through metallic lines and $R_{23}$ decreases. The calibration from \citet{2004ApJ...613..898T} is valid only for the upper branch of the $R_{23}$ vs $12+log{O/H}$ relation.  We have assumed that our sample of intermediate-mass galaxies stands in the upper branch. In fact, galaxies in the extreme end of the lower branch are extremely rare and are associated with dwarf galaxies. For moderate metallicities, near the turning point of the relation, the uncertainties in selecting the appropriate branch is smaller than the uncertainties from sky and extinction.  We test this hypothesis with the [\ion{N}{ii}]/[\ion{O}{ii}] indicator. There are only 4 galaxies in our sample with [\ion{N}{ii}] measurements: all have $\log{f([\ion{N}{ii}])/f([\ion{O}{ii}])}> -1$ and therefore belong to the upper-branch \citep{2002ApJS..142...35K}. We think that, with a large level of confidence, all the objects stand in the upper branch of the $R_{23}$ vs $12+log(O/H)$. Nine objects have $\log{R_{23}}> 1$ where the \citet{2004ApJ...613..898T} calibration is not defined. In such case we have adopted the limit given where $\log{R_{23}}= 1$. 

For the 5 lowest redshift galaxies (type E3 in Table 2), the [\ion{O}{ii}] emission line falls out the wavelength range. We have used in this case the $R_3$ parameter defined by $R_3$=1.35$\times$([\ion{O}{iii}]$\lambda$5007/$\mathrm{H\beta}$) \citep{1984MNRAS.211..507E}. The oxygen abundance have been estimated with the calibration proposed by \cite{1992ApJ...401..543V}:
\begin{equation}
12+log(O/H)=-0.60\times\log{R_3}-3.24 \, ,
\end{equation}
We have not found any evidence of systematic bias between the $R_{23}$ and the $R_{3}$ calibrations.
Metal abundance and error associated are shown in Table \ref{table3}. 

\subsection{ Error budget }
The uncertainties in the data were assumed to be Poisson distributed. The error budget of emission line flux includes contributions from the sky and the object. We have propagated this error on the $\mathrm{H\gamma/H\beta}$ line ratio to estimate the error on $A_{\mathrm{V}}$(Balmer). For $A_{\mathrm{V}}$(IR) the error has been estimated taking into account the error on the IR luminosity and the Poisson noise of the $\mathrm{H\beta}$ line. As the error on $A_{\mathrm{V}}$(Balmer) is dominated by the error on the $H_{\mathrm{\gamma}}$ line, we can assume that the errors on $A_{\mathrm{V}}$(Balmer) and  $A_{\mathrm{V}}$(IR) are not correlated. Thus, the error of the final extinction depends on the error of the two extinction estimates.  The errors of line ratios and metallicity have been calculated by Monte-Carlo simulation taking into account the error on extinction and the sky error on lines flux. 

\section{Stellar mass - metallicity relation and its evolution}
\subsection{The Stellar mass - metallicity relation at $\mathrm{z}\sim0.7$}
\begin{figure} \resizebox{\hsize}{!}{\includegraphics{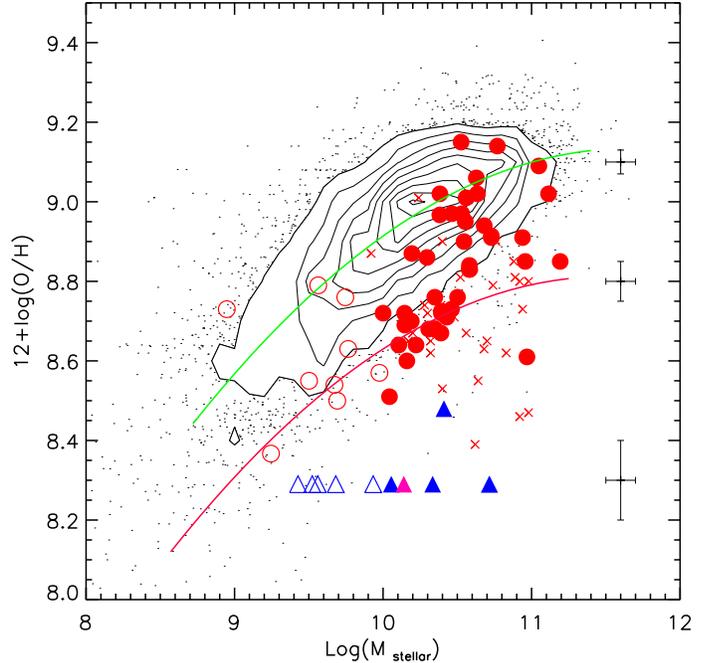}} \caption{Stellar mass-metallicity relation for the SDSS galaxies in contours and dots (z=0, \citet{2004ApJ...613..898T}, \citet{2003ApJS..149..289B} see text ) and the 88 intermediate-z galaxies: sample A in red circle and sample B in red cross. Galaxies with masses under the completeness limit $\log{M_{\mathrm{stellar}}}< 10$ are plotted with open symbols. The objects in sample~A with $\log{R_{23}}>1$ are plotted as small pink triangle and those detected as possible AGN are plotted as blue triangles. The typical error bars for bin~1, bin~2 and bin~3 are plotted in the right. The red line indicates the median of the intermediate-z relation.The green line represents the local relation fit found by \citet{2004ApJ...613..898T}. The shift between the local relation and our data sample is $\Delta$[12+log(O/H)]= -0.31\,dex$\pm$0.03. } \label{M-Z1} \end{figure} 

We compared the metal abundance of our sample of 88 distant galaxies with those of local starbursts from SDSS \citep{2004ApJ...613..898T}. Stellar masses have been estimated using absolute K band magnitude with $M_{\mathrm{stellar}}/L_{\mathrm{K}}$ expected by observed rest frame B-V color  \citep{2003ApJS..149..289B} and converted to a \citet{1955ApJ...121..161S} IMF. The convertion between \citet{2001MNRAS.322..231K} and \citet{1955ApJ...121..161S} is described in \citet{2003ApJS..149..289B}. We have found that distant galaxies are metal deficient compared to local starbursts, as shown in Fig. \ref{M-Z1}. Given the small range in stellar mass covered by our sample, it was not possible to constrain the evolution of the shape of the M-Z relation at z$\sim$0.7. Thereafter, we have assumed that the slope of the M-Z relation has remained unchanged compared to the local relation. At z$\sim$0.7 the local relation found by \citet{2004ApJ...613..898T} is shifted to lower metallicity by $\Delta$[12+log(O/H)]=0.31\,dex$\pm$0.03. For each galaxy we have calculated the difference between its metallicity and the metallicity given by the \citet{2004ApJ...613..898T} relation for its stellar mass. The offset of the z$\sim$0.7 relation is the median of this difference. Only galaxies with a stellar mass over the limit of completeness have been used. The standard error has been estimated using a bootstrap method. 

The shift is assumed to be only due to a metallicity evolution. Indeed, \citet{2006A&A...447..113L} argue that the evolution in the M-Z relation is due to a diminution of the metal content in galaxies rather than an increase of their stellar mass. In fact, the amount of star formation necessary to increase by ten times the stellar mass would imply that almost all the light in distant galaxies is associated with very young stellar population. However, the spectra of these galaxies show a mixture of young, intermediate and old stellar populations. 

There is a large dispersion of the data around the median relation at z$\sim$0.7, about $\pm$0.45\,dex. To verify if this dispersion is intrinsic to the objects or due to the error on our data, we have divided our sample into three bins. Bin 2 (3) includes the 25\% of object which are the most over (under) metal abundant compared to the median. The rest of the objects constitutes bin 1. We have calculated the median error of the metallicity in each bin and found : 0.05, 0.03 and 0.1 (bin 1, bin 2 and bin 3). We have plotted the mean error of each bin in Fig.7. There is an evolution of the error of the metallicity through the 3 bins, which is well explained by the shape of the 12+log(O/H) vs $R_{23}$ calibration. Indeed, the metallicity varies rapidly with $R_{23}$ in metallicity range near the turn over of the function, at low metallicities.  Nevertheless, the dispersion of the data in the M-Z relation is much larger that the error on metallicity. The dispersion is then intrinsic to the objects. We discuss about the origin of this dispersion in Sect. 6.4.

We have searched for systematic effects due to morphological and environment properties. First we have isolated the sub-sample of galaxies overstanding within in the CDFS structures. Metal abundance may be higher in these galaxies because of higher merger rate and then rapid galaxy evolution. Nevertheless we have not found any evidence of bias. This is probably due to an effect of data selection. The working sample majority includes starburst galaxies with intermediate mass. The evolved galaxies in structures are expected to be massive galaxy with poor star-formation. 

We have also tested the effects of morphology in the locus of galaxies in the M-Z relation. We have selected spiral galaxies in sample~A+B \citep{2005A&A...435..507Z} and have plotted them in the M-Z plot, but no effect has been found. 

Finally, we have focused particularly on the study of LIRGs. Recently, \citet{2008ApJ...674..172R} found that LIRGs and ULIRGs are metal under-abundant by a factor two compared to emission-line galaxies in the local Universe. We have looked for an evidence in the sub-sample of 20 LIRGs representing 20\% of the sample~A+B. We have compared the LIRG metallicity with the starburst galaxies at z$\sim$0.7 and we did not find any shift. The small metal overabundance found by \citet{2006A&A...447..113L} was due to a selection effect: LIRGs in sample~B are more massive and have higher redshift than in sample~A. Nevertheless, we have noticed that the z$\sim$0.7 LIRGs fall in the same locus as the local LIRGs in the M-Z relation. We will discuss these results in the section 6.

\subsection{Evolution to high redshift of the stellar mass - metallicity relation}
\begin{figure*} \centering \includegraphics[width=17cm]{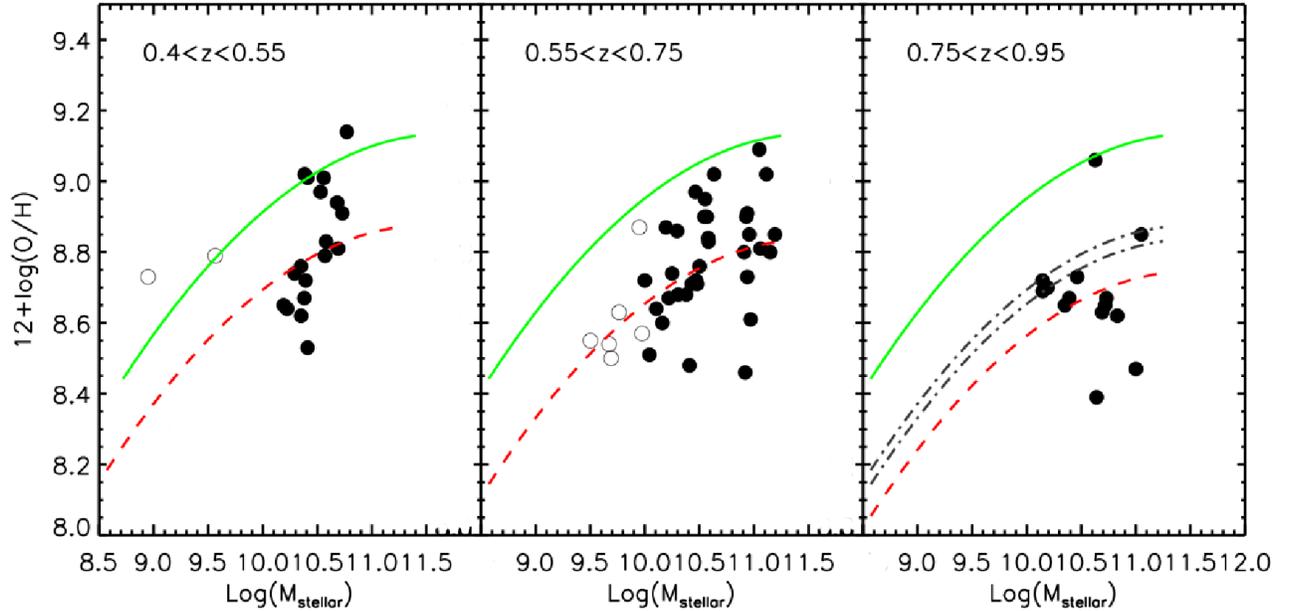} \caption{\textbf{Evolution of the M-Z relation inside the Sample~A+B}. \textbf{Left Panel}:The 17 galaxies from the $0.4\leq z<0.55$ bin. \textbf{Middle Panel}:The 38 galaxies from the $0.55\leq z<0.75$ bin. \textbf{Right Panel}: The 15 galaxies from the $0.75\leq z\leq 0.95$ bin. The galaxies with masses under the completeness limit are plotted with open circles. Objects with $log(R23)>1$ and AGN are not plotted. The median of each bin is marked in dashed red line, see table \ref{offset_evo} for the $\Delta$[12+log(O/H)]. The local relation from \citet{2004ApJ...613..898T} is marked with a green line. Only galaxies with over the completeness limit $\log{M_{\mathrm{stellar}}}< 10$ have been taken to calculate the median shift. The two black lines in the right panel are the median of the two previous bin.} \label{M-Z_multi} \end{figure*}

One of the aims of our work is to robustly establish the evolution of the M-Z relation. We have first searched for a redshift evolution of the metal content in galaxies inside the sample of intermediate redshift galaxies. As sample A+B covers a large range in redshift, from z=0.4 to z=0.95, we have split it into 3 redshift bins: $0.4<z<0.55$, $0.55<z<0.75$ and $0.75<z<0.95$. In each redshift bin, we have measured the median shift of the local relation toward lower metallicity. The error of the median offset has been calculated by a bootstrap method. Fig. \ref{M-Z_multi} shows the evolution of the M-Z relation along the 3 redshift bins. The median offsets from the local relation in the 3 bins are given in table \ref{offset_evo}. We find an evolution of the M-Z relation from z$\sim$0.45 to z$\sim$0.85 of $\Delta$[12+log(O/H)]=$-$0.13\,dex.
\begin{table} 
\caption{Offset between local relation and the 3 redshift bin.} 
\label{offset_evo} 
\centering
\begin{tabular}{c c c} 
\hline
Redshift bin & $\Delta [12+log(O/H)]$& $ error \Delta [12+log(O/H)]$ \\ 
\hline 
$0.40\leq z<0.55$ & -0.26 & 0.08 \\
$0.55\leq z<0.75$ & -0.30 & 0.04 \\
$0.75 \leq z \leq 0.95$ & -0.39 & 0.09 \\
\hline 
\end{tabular} 
\end{table} 

We have then searched for an evolution of the oxygen abundance at higher redshifts. It is important to notice that the comparison between different samples needs consistent measurements of oxygen abundance and stellar mass. In fact, particular care has to be taken when comparing metallicities from different metal estimators. Especially when the aim is to evaluate the offset between samples due to a evolution in metallicity. Remember that the systematic error due to the use of different calibration can range up to 0.5\,dex and thus can be higher than the metal evolution, cf. \citet{2008ApJ...674..172R}.
We describe below the high-z samples from the literature which have been used for this comparison.
\begin{itemize}
\item \citet{2008ApJ...678..758L} have recently published the mass-metallicity relation for 20 galaxies from the Deep 2 Galaxy Redshift Survey at $1.0<z<1.5$. We have only considered the 7 galaxy spectra having individual oxygen mesurements. As only [\ion{O}{iii}], $\mathrm{H\beta}$, [\ion{N}{ii}] and $\mathrm{H_\alpha}$ emission line measurements were available from the NIRSPEC/Keck observations, the $N2$ parameter has been used with the \citet{2004MNRAS.348L..59P} calibrator to estimate metallicity. The stellar masses were calculated from K-band photometry and the procedure proposed by \citet{2005ApJ...625..621B}.
\item At higher redshift, z$\sim$2, \citet{2004ApJ...612..108S} observed 7 star-forming galaxies. The oxygen abundance has been also estimated using the $N2$ parameter and the \citet{2004MNRAS.348L..59P} calibrator.  The stellar masses have been calculated by fitting the spectral energy distribution to multi-band photometry. 
\item At the same redshift, $\langle z\rangle$=2.26, \citet{2006ApJ...644..813E} have measured the metal abundance of 6 composite spectra from 87 star-forming galaxies selected by ultraviolet rest-frame and binned by stellar mass. They have used  the same calibration for $12+log(O/H)$ estimation and used the same stellar mass estimates as \citet{2004ApJ...612..108S}.
\item \citet{2008arXiv0806.2410M} have recently presented preliminary results from the AMAZE large program which aim to constrain the M-Z relation at z=3. The oxygen abundance has been measured by the $R_{23}$ parameter and following the procedure described by \citet{2006A&A...459...85N}. We have only selected the 5 objects above our stellar mass completeness limit and have used the $R_{23}$ measurement to estimate the metal abundance using the \citet{2004ApJ...613..898T} calibration. 
\end{itemize}

Three samples use the $N2$ parameter to evaluate the oxygen abundance. Unfortunately, metallicity estimation based on $N2$ and $R_{23}$ can give strongly discrepant results and thus cannot be directly compared. The $N2$ parameter has several disadvantages: the [\ion{N}{ii}]/$\mathrm{H_\alpha}$ ratio saturates at $12+log(O/H)\geq8.8$ and becomes insensitive the metal abundance. Furthermore, it is very sensitive to the ionization parameter and the contribution of the primary and second nitrogen whose evolution is not yet well constrained. To compare the galaxy metallicities using the $N2$ parameter we have decided to re-evaluate the metallicities using the \citet{2002MNRAS.330...69D} calibration. Indeed, this calibration is similar to the \citet{2004ApJ...613..898T} and other $R_{23}$ based calibration in the working range $8.4\leq12+log(O/H)\geq8.7$ \citep{2008arXiv0801.1849K,2004ApJ...613..898T}. For the galaxies in the sample A+B having [\ion{N}{ii}] measurements, we have compared the metallicity to those of several $N2$ calibrations. We have confirmed that the \citet{2002MNRAS.330...69D} calibration give similar results than the abundances estimated in this work. The M-Z diagram for the local, intermediate and high redshift samples are plotted in Fig. 9. 

We have measured the median offset of the local relation and the associated error in all the high-z samples. The offset $\Delta$(O/H) for the 3 redshift bins of the intermediate galaxies and for the four high-z samples are plotted in Fig. \ref{lookback} as a function of the lookback time. We found that the evolution of the mean metallicities from local Universe to a lookback time of $\sim$12\,Gyrs is linear and with a slope given by:
\begin{equation}
\log{Z_{\mathrm{tb}}/Z_0} = -0.046\times t_{\mathrm{Gyrs}} \,,
\end{equation}
If the \citet{2004MNRAS.348L..59P} is used for metallicity estimate of high-z samples, the evolution remains linear with a slope of $-$0.051. 
For comparison we overplot the metallicity evolution extrapolated by \citet{2005ApJ...635..260S}. The empirical evolution model has been constructed by shifting the local M-Z relation in $\log{M_{\mathrm{stellar}}}$ contrary to current work where the shift has been evaluated in term of metallicity for the reason explained above. Their empirical model is concordant with a close box model with an exponential star formation efficiency. We discuss in sections 5 and 6 the discrepancy between our observations and those of \citet{2005ApJ...635..260S}, as well as the consequences for our understanding of the evolution of the M-Z relation.

\begin{figure} \resizebox{\hsize}{!}{\includegraphics{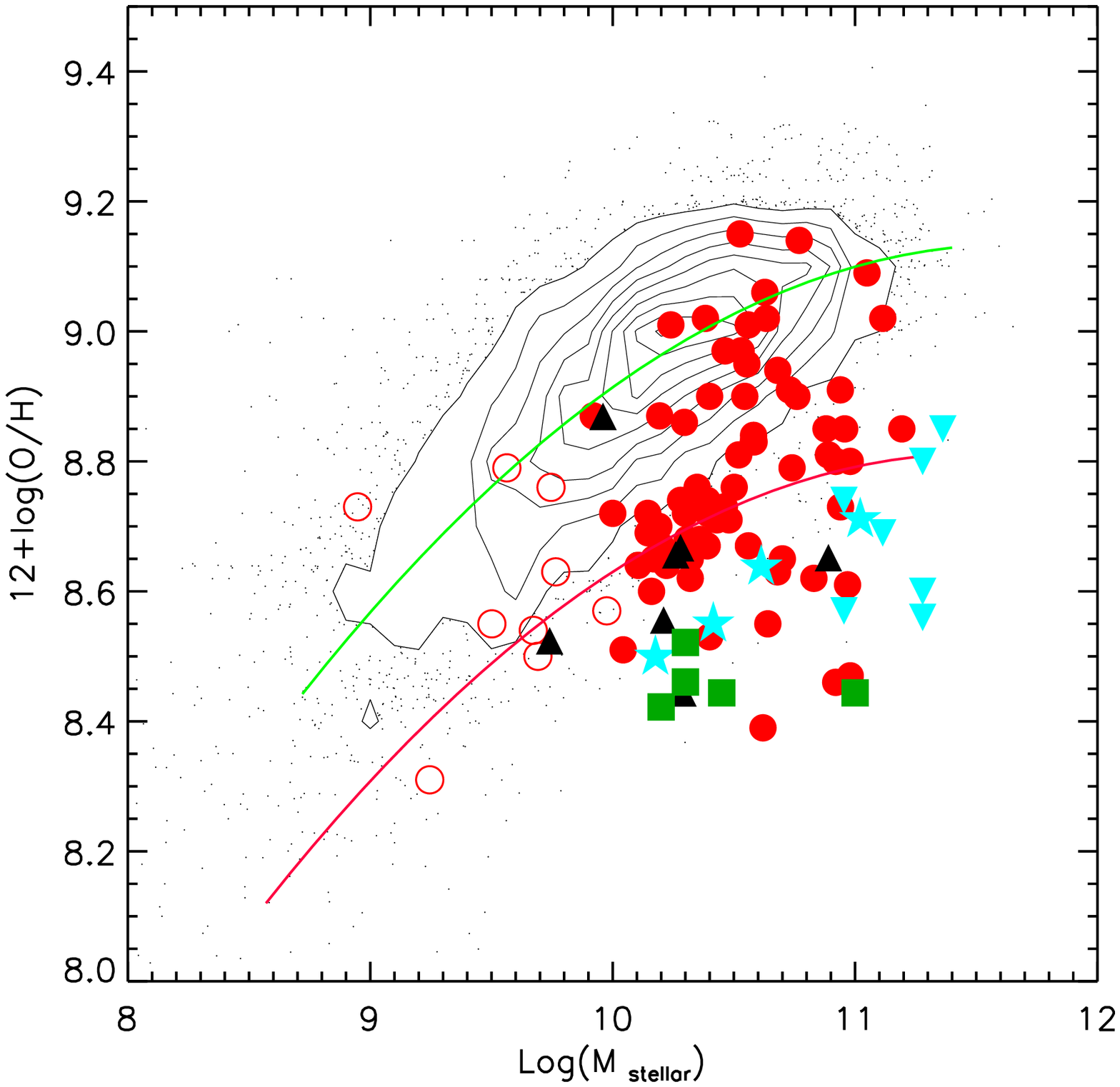}} \caption{Stellar mass-metallicity relation for the SDSS galaxies in contours and dots (z=0, \citep{2004ApJ...613..898T, 2003ApJS..149..289B} ) and the 58 intermediate-z galaxies above the completeness mass limit in red fill circle. Galaxies under the completeness mass limit are plotted in red open circle. The local relation is plotted in green \citep{2004ApJ...613..898T} and the median of the intermediate-z relation is in red. The 7 individual galaxies at z$>$1 of \citet{2008ApJ...678..758L} are plotted as black triangle. The two z$>$2 samples are plotted in blue: inverted blue triangles indicate the 7 star-forming galaxies of \citet{2004ApJ...612..108S}; blue stars indicate the 5 stellar mass bin of \citet{2006ApJ...644..813E}. The $\mathrm{z}>3$ sample of Maiolino et al. (2008) are plotted in green squares. } \label{M-Z_highz} \end{figure} 

\begin{figure} \resizebox{\hsize}{!}{\includegraphics{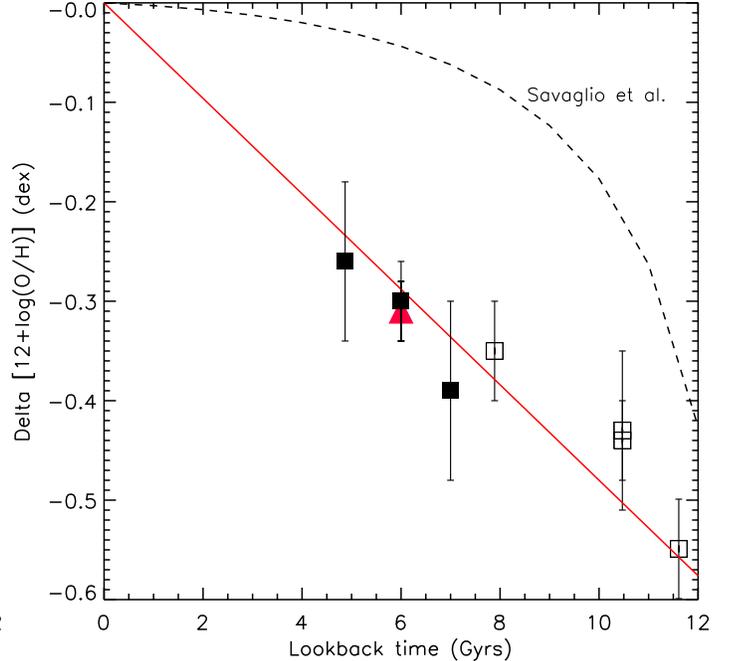}} \caption{ The metallicity shift from the local relation of the 4 high-z sample as a function of the lookback time is plotted as black open squares. The metallicity shift for the 3 redshift bin of the IMAGES sample are in black squares and the median of the 3 bins as a red triangle. The red line is the linear fit of the 5 high-z data points verifying $\Delta$[12+log(O/H)]=0 at z=0. The evolution found by \citet{2005ApJ...635..260S} in the frame of a close box model is indicated with a dashed line for $\log{M_{\mathrm{stellar}}}$=10.45 galaxies.} \label{lookback} \end{figure}

\section{Possible systematic uncertainties}
In this section we discuss the possible systematic errors that can affect our determination of the evolution of the M-Z relation. 
\subsection{Metallicity calibrators}
As we have emphasized in \S 3, the determination of metallicity using different calibrations may give very discrepant results. However, \citet{2008arXiv0801.1849K} have shown that the relative metallicities between galaxies can be reliable estimated within $\sim$0.15\,dex when using the same strong line calibration. In this work we have taken a special care to compare our sample to the local galaxies using the same metallicity calibration and methodology. \citet{2008ApJ...678..758L} have compared the M-Z relation at z$\sim$1 to the local relation using $N2$ for both redshifts. They have found that z$\sim$1 galaxies have metal abundance 0.22\,dex lower than in present Universe. This is quite comparable to our results. 

The comparison between the metallicity at z$\sim$0.65 and higher redshift samples is less robust: low S/N spectra and different metal indicators have been used. In fact, high-z samples use usually the $N2$ indicator. Several authors have tried to build conversions between metallicity calibrations, e.g. \citet{2008arXiv0801.1849K}. However, these conversions have been derived from local samples and may be not valid at higher redshift. \citet{2008ApJ...678..758L} have suggested that at z$\sim$1 the metallicities of galaxies are overestimated by $\sim$0.16\,dex due to different physical conditions in HII regions at higher redshift: violent star-formation, shocks, higher ionization parameter. No evidence of such different conditions of HII regions have been found in our sample, according to the diagnostic diagrams.  

\subsection{Aperture effect}
The effect of the aperture when comparing intermediate redshift data to the SDSS sample has been widely studied by several authors \citep{2005ApJ...635..260S, 2004A&A...423..867L, 2004ApJ...617..240K,2008arXiv0801.1849K}. The SDSS observed only the center part of galaxies with a $3''$ aperture fiber. The metallicities estimated are expected to be higher than the average metallicity of the galaxies because of the metallicity gradient with the distance from the center. \citep{1992MNRAS.259..121V,1994ApJ...420...87Z,1997ApJ...489...63G,2000A&A...363..537R}. \citet{2008arXiv0801.1849K} have shown that the effect is relevant for massive galaxies $M_{\mathrm{stellar}}>10^{10}\mathrm{M_\odot}$. However, the bias due to aperture effect in the SDSS is not larger than Z$\sim0.1\,\mathrm{\mathrm{dex}}$ \citep{2005PASP..117..227K} and thus have a minor impact on the estimation of the evolution  of the M-Z relation. 

\subsection{Extinction}
The effect of extinction on the estimation of metallicity is one of the main systematic errors affecting the M-Z relation. For example, an underestimation of 1.5\,mag in extinction leads to an overestimation of $\sim$0.15\,dex in oxygen abundance. Several works have assumed an average extinction to correct spectra of galaxy delimiting a large range in stellar mass. However, there is a correlation between mass and extinction \citep{2007A&A...473..411L}: massive galaxies have higher extinction. Assuming a median extinction leads to an overestimation of the metallicity in massive galaxies and an underestimation in low mass system. At z$\sim$0.6, LIRGs which have high content of dust and gas represent 15\% of intermediate mass galaxy population. Extinction is expected to have a large impact on any abundance estimate at this redshift. 

The effect of extinction may be one of the reasons for the difference between our observations and those of \citet{2005ApJ...635..260S}. They have observed that the average metallicity of massive galaxies has not evolved from z$\sim$0.6 to present days and thus have claimed evidence for downsizing in metallicity. However, the majority of massive galaxies in the \citet{2005ApJ...635..260S} sample come from measurements of \citet{2003ApJ...597..730L} which are from spectra with low S/N and spectral resolution. Moreover, they have applied an average $A_V=1$ to correct emission lines. This extinction value is underestimated by about $\sim$1\,mag for at least one third of the \citet{2003ApJ...597..730L} sample, which corresponds to LIRGs. \citet{2006A&A...447..113L} have estimated that the underestimation of the extinction in the \citet{2003ApJ...597..730L} sample leads to an overestimation of $\sim$0.3\,dex on metallicity, which can explain the discrepancy with our sample. The galaxies in the GDDS sample in \citet{2005ApJ...635..260S} have low luminosities and can be also affected by the low S/N of the spectra which does not allow to retrieve proper extinction from Balmer lines for individual galaxies. These effects are amplified for evolved massive galaxies especially those experiencing successive bursts and containing a substantial fraction of A and F stars. Unfortunately these stars are predominant in intermediate mass galaxies, and thus contaminate extinction measurement by their large Balmer absorption. Furthermore, the average extinction used by \citet{2005ApJ...635..260S} can be underestimated because the combining spectra used to measure $\mathrm{H\gamma}$ and $\mathrm{H\beta}$ lines is dominated by low extinction systems.

\subsection{Stellar mass}
A bias in the estimation of stellar mass at high redshift must be considered. In this work, stellar masses have been estimated from $M_{\mathrm{stellar}}/L_\mathrm{K}$ ratio using the method of \citet{2003ApJS..149..289B}. This method uses the tigh correlation between rest frame optical color and $M_{\mathrm{stellar}}/L_\mathrm{K}$ ratios. The amount of light due to red giant stars is corrected in the $M_{\mathrm{stellar}}/L_\mathrm{K}$ using the g-r colors. At higher redshift, the influence of TP-AGB stars in the derivation of the stellar masses could result in an overestimation of the stellar mass by $\sim$0.14\,dex \citep{2006ApJ...652...85M,2007A&A...474..443P}. However this bias has only a thin effect on the evolution of the M-Z relation. The bias in stellar mass directly translated into a systematic effect of the evolution of the M-Z relation is around -0.05\,dex.

\subsection{Uncertainties budget}
We summarize all possible sources of systematic uncertainty identified so far in Tables 4 and 5. We expressed all of them in terms of their influence in the shift of the M-Z relation from local Universe to higher redshift. In the intermediate redshift sample we have taken a particular care to diminish the bias from extinction estimates and from metallicity calibrations. In the first place, the methodology has allowed us to reduce the bias due to extinction. Secondly, we have used a similar methodology and the same metallicity calibrators for our sample and the local sample of \citet{2004ApJ...613..898T}. The remaining systematics may come from stellar mass estimates and from the aperture effect on the SDSS data. However, as illustrated in Table 4, the systematic error from these two sources is small and thus does not change our results. 

The shift of the M-Z relation from intermediate to high redshift is less secure. The sources of systematic error come from the use of different metallicity calibrations and a possible evolution of the HII properties at high-z, see Table 5. We have tried to minimize the bias from metallicity estimates selecting for the high-z sample, the metallicity calibration that better follows the one used in this work. However, even using the \citet{2004MNRAS.348L..59P} calibration instead of the \citet{2002MNRAS.330...69D} calibration, we do find a linear evolution of the metal content in galaxies and with a similar slope. As the evolution of the M-Z relation is strongly constrained at z$\sim$0.6, the bias in the high-z sample will not change the shape of the evolution function but only the slope. Even accounting for all the systematics shown in Table 5. the observations from intermediate and high redshift samples are not compatible with the evolution found by \citet{2005ApJ...635..260S}.
  
\begin{table} 
\caption{Identified systematic uncertainties that could impact on the shift of the M-Z relation between local and z$\sim$0.6 galaxies. } 
\label{budget} 
\centering
\begin{tabular}{l c c} 
\hline
Possible bias & $ Z$ & Comments \\ 
\hline 
Aperture effect &0.1& See Sect. 5.2 \\
Stellar mass&-0.05&  See Sect. 5.4\\
\hline 
\end{tabular} 
\end{table} 

\section{Discussion}
\subsection{Discarding the closed-box model}
We first compare our observations with a simple model of galaxy evolution: the closed-box model \citep{1972ApJ...173...25S,1980A&A....89..246T}. In this model galaxies are isolated systems which evolve passively transforming their gas into stars, enriching their content in metals. The amount of gas converted into stars from z$\sim$0.6 to present can be estimated, following the same methodology as \citet{2006A&A...447..113L} and proposed by \citet{2004ApJ...617..240K}. The metal abundance Z is related to the gas mass fraction by
\begin{equation}
Z(t)=Y\,\ln{1/\mu(t)} \, ,
\end{equation}
where $\mu=M_{\mathrm{gas}}/(M_{\mathrm{gas}}+M_{\mathrm{stellar}})$ is the fraction of gas. As the yield is constant the variation of metallicity depends only on the variation of the gas fraction :
\begin{equation}
\mathrm{d}(\log{Z})/\mathrm{d}\mu=0.434/\mu\mathrm{\ln}{\mu} \, ,
\end{equation}
We assume that local galaxies have a gas fraction $\mu_0$=10\%. The metallicity of the gas phase has decrease by two from z=0 to z=0.6. Then by equation (7) the gas fraction at z=0.6 predicted by the close box model is around mu=30\%. This means that the present day intermediate galaxies would have converted 20\% of their present baryonic mass, from gas into stars. Recently, \citet{2008A&A...484..173P} have studied the evolution of the Tully-Fisher relation at z$\sim$0.6 using a representative sample of 65 emissions line intermediate galaxies. They have found a significant evolution of the K-band TFR zero point, which they attribute to an average brightening by 0.66\,mag. After considering several alternatives for the evolution of the stellar mass, they conclude that on average, spiral galaxies have doubled their stellar masses since z=0.6. The closed-box model predicts a much lower efficiency in converting gas into stars during the same elapsed time. There would be not enough gas from the close box model to feed the stellar mass doubling. This discrepancy can be explained if $\sim$30\% of the stellar mass was formed by external gas supply probably metal enriched \citep{1990MNRAS.246..678E}. We can definitively rule out a closed-box model as a viable description of the metal content in galaxies from z=0.6 to z=0. 

The closed-box model has been taken as a first approach to explain the metal evolution of galaxies, e.g.\citet{2005ApJ...635..260S}. However, we notice that the close box model is not consistent with the two major scenarios of galaxy evolution. Indeed, to resolve the G-dwarf problem and to explain the evolution of disk, the secular scenario needs to take into account large amount of infalling gas, e.g. primordial gas in filaments \citep{2005A&A...441...55S}. In the hierarchical scenario the evolution of galaxies is driven by mergers and interactions, thus by gas exchanges.
  
\subsection{Comparison with previous models of galaxy chemical evolution}
We compare our results with galaxy evolution simulations by \citet{2007MNRAS.374..323D, 2008A&A...483..107B, 2008arXiv0801.2476M}. We have only compared the evolution of metal content in the gas phase founded by the models in the same mass range than in our sample. 

Within the framework of the hierarchical scenario, \citet{2007MNRAS.374..323D} have simulated the evolution of the M-Z relation from present to z=3. They predict very little evolution on the relation of only $-$0.05\,dex since z=3 and that massive galaxies with $M_{\mathrm{stellar}} > 3\times10^{10}\,\mathrm{M_\odot}$ have reached the local relation well before z=1. Thus they predict that the spectra of massive galaxies should be mainly composed of an old metal-rich stellar population. This is in strong contradiction with our results. In our study 38\% of the galaxies are above the critical mass and have strongly increased their metal content, which is consistent with the evolution of the Tully Fisher relation \citep{2008A&A...484..173P}. Moreover, the results of stellar continuum fitting indicates that the predominant stellar population is composed of A and F stars, which suggest a significant stellar formation 1.5\,Gyr prior to our observations. The predominance of intermediate age stellar population in galaxies at intermediate redshifts had been previously pointed out by \citet{2001ApJ...550..570H}. 

\citet{2008arXiv0801.2476M} have simulated numerical models of galaxy evolution in the hierarchical scenario and they also predict an important downsizing effect. Metal abundances of galaxies with $M_{\mathrm{stellar}} > 10^{10}\,\mathrm{M_\odot}$ do not evolve since z$\sim$1.2. However, they have found a stellar mass assembly and a metallicity dispersion consistent with our results. 

Finally we compared our observations with the chemical evolution model presented by \citet{2008A&A...483..107B}. They investigate the star formation history of star-forming galaxies from z=0 to z=1 and compare with evolution models developed for local spiral galaxies within the frame of a secular evolutionary scenario. From a model they predict the SFR and the M-Z relation evolution for intermediate mass galaxies. Their predicted evolution of metallicity is half of our observations: they predict a shift of  $\sim-$0.2\,dex at z=1 (against $\sim-$0.4\,dex in this study, see Table 3). 

They also predict that at z$\sim$1 massive galaxies have increased more rapidly their metal content than low mass galaxies which is at odds with most current models. The \citet{2008A&A...483..107B} model is based on the evolutionary history of the Milky Way, a galaxy which may not be representative of typical spirals \citep{2007ApJ...662..322H}. For example, it is unlikely that the Milky Way has experimented in the past 8\,Gyrs a strong episode of star-formation such as a LIRG, which is quite common in galaxies of our sample. 

At z=0.4-0.9, the fraction of emission line galaxies is 60-70\% (e.g. \citet{1997ApJ...481...49H}). As such, any deviations from local values in a sample of emission line galaxies is at odd with the above modelling. 
If we constrain Sample~A to the redshift range from 0.4 to 0.7 (limit technical problems like lines out of range), we find that 96/136 galaxies are emission lines galaxies (31 with EQW(OII)$<$15 and 65 with EQW(OII)$>$15) and only 40 are absorption line galaxies. From the 65 emission lines galaxies with EQW(OII)$>$15, 40 have been studied Ðthe rest being rejected mainly for instrumental reasons (sky lines, instrument fakes etcÉ).  

Thus existing modeling have difficulties in reproducing our observations, i.e. a strong evolution of the metal abundance of intermediate mass galaxies over the last 6\,Gyrs.

\subsection{The star-formation efficiency of intermediate mass galaxies at z$\sim$0.6}
We estimate the doubling time, defined by $t_{\mathrm{D}}=M_{\mathrm{stellar}}/SFR_{\mathrm{IR}}$, for emission line galaxies in the IMAGES sample. The SFR for galaxies without IR detection have been estimated using the $L_{\mathrm{IR}}$ limit of detection of MIPS/Spitzer and the $SFR_{\mathrm{2800\AA}}$. To determine the unobscured $SFR_{\mathrm{2800\AA}}$, we used the UV calibration of \citet{1998ARA&A..36..189K} and the rest-frame 2800$\AA$ luminosity from SED fitting.

We confirm that massive galaxies convert less gas into stars than lower mass galaxies at $0.5<z<1$: e.g the $t_{\mathrm{D}}$ for a massive galaxy $\log{M_{\mathrm{stellar}}}$=11 is above 3\,Gyrs whereas a less massive galaxy log$M_\mathrm{\mathrm{stellar}}$=10.3 doubles its stellar content in less of 1\,Gyrs. This effect is usually called downsizing: low mass galaxies are forming stars at latter epoch than massive galaxies. Severals authors have pointed out that downsizing is in contradiction with the hierarchical scenario. Massive galaxies form later than low mass galaxies by accretion of small dark halos, in the bottom-up scenario. However \citet{2006A&A...459..371M} and \citet{2006MNRAS.372..933N} have demonstrated that the downsizing of star-forming galaxies is inherent in gravitational processes of the hierarchical scenario. Low mass galaxies have very long gravitational collapse time-scales due to their small potential well. Thus they start to produce stars later than massive galaxies.

The small values of doubling time in intermediate mass galaxies at z$\sim$0.6 (see Figure 12) confirm the strong evolution of these sources at this epoch. We do find that the doubling time diminishes towards redshift which is due to the star-formation increase with redshift.These observations are in good agreement with the results of \citet{2008A&A...484..173P} who argue that galaxies have doubled their stellar mass in the last 6\,Gyrs. 

\begin{figure} \resizebox{\hsize}{!}{\includegraphics{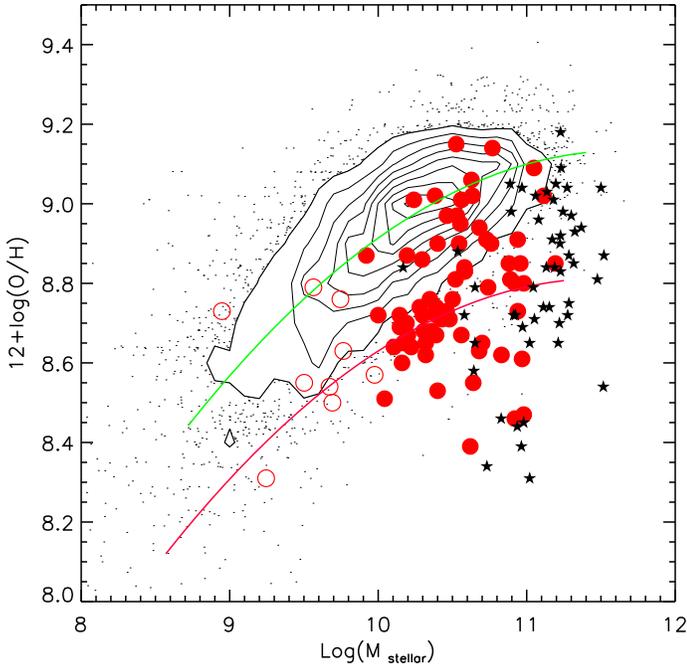}} \caption{M-Z relation for starburst galaxies at z$\sim$0.6 and local LIRGs from \citet{2008ApJ...674..172R}. The symbols used in this plot are the same as in fig. 9. The local LIRGs are plotted as black stars. } \label{double_z} \end{figure} 

\subsection{Similarities between local LIRGs and distant starbursts \& LIRGs}
The IMAGES sample has enabled us to investigate several properties of distant starbursts, such as metallicity and star-formation rates, etc. These observations reveal that local LIRGs and distant starbursts/LIRGs share many properties: 
\begin{itemize}
\item \textbf{Same locus in the M-Z relation}\\
\citet{2008ApJ...674..172R} have shown that local LIRGs do not follow the local M-Z relation. Recall that the average SDSS galaxies populating the local M-Z plane are forming stars at moderate rate (less than few solar masses per year), much lower than for starbursts and LIRGs. Local LIRGs (see Figure 11) have a deficiency in metal compared to other local galaxies and delineate the same locus in the M-Z relation than the high-z starbursts \& LIRGs.  Both populations show a large and similar dispersion of their metallicity at a given stellar mass.  Thus local LIRGs and distant starbursts share a same metal content at a given stellar mass. Notice that in this work, we have not found differences between distant LIRGs and distant starbursts.

\item \textbf{Similar high star-formation efficiency}\\
Fig. 12 shows the distribution of doubling times for local LIRGs \citep{2008ApJ...674..172R} and IMAGES galaxies. In order to properly compare the two samples, we used a same stellar mass range to avoid biases (e.g. lower doubling time from low mass galaxies in the intermediate mass galaxy sample). The two distributions are very similar. The Kolmogorov-Smirnov test attests that the probability that the two distributions arise from the same population is about 40\%. We notice that stellar mass for local LIRGs has been estimated from their absolute magnitude $M_{K}$ and assuming a $M/L_{\mathrm{K}}$=1. This crude assumption can lead to a slight overestimation of their stellar mass (by $\sim$0.1\,dex) although it does not affect the overall result. 

\begin{figure} \resizebox{\hsize}{!}{\includegraphics{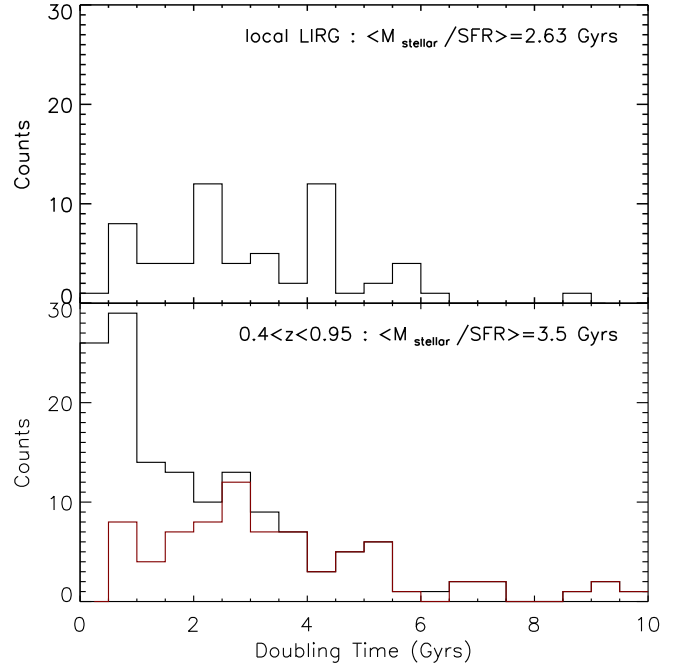}} \caption{Histogram of the doubling time of local LIRGs from \citet{2008ApJ...674..172R} (upper panel) and emission lines galaxies from the IMAGES sample (botton panel). 
The distribution of IMAGES sample for galaxies delimiting the same $\log{M_{\mathrm{stellar}}}$ than local LIRGS is overplotted in red.} \label{double_hist} \end{figure} 

\item \textbf{Similar morphologies: dominated by mergers and spirals} \\
Severals studies in the local Universe have pointed out the link between LIRGs, merging of gas-rich spirals\citep{2002ApJS..143..315V, 2004ASPC..320..230S,2006ApJ...649..722W}. At higher redshifts a significant fraction of the distant galaxies have morphologies that include spiral to peculiar \citep{2005A&A...435..507Z,2008A&A...484..159N}. Distant LIRGs are also associated with merger at different stages or with large disks \citep{2005A&A...435..507Z, 1999ApJ...517..148F}.

\end{itemize}
Fundamental properties such as morphologies, star-formation efficiency and mass-metallicity are similar for both local LIRGs and distant starburst/LIRGs. We may thus wonder whether local LIRGs and distant starburst also share the same physical mechanisms explaining their properties. The only discrepancy between the two populations is their numerical abundance: while local LIRGs only represent 0.5\% of the local massive galaxies, distant starbursts studied here represent approximatively 60\% of distant massive galaxies \citep{1997ApJ...481...49H}. We suggest that local LIRGs are the latecomers of the main population of massive starburst galaxies visible at z$\sim$0.6. 

What is the physical process explaining the properties of these two galaxy populations of actively star forming galaxies? Since star formation increases the abundance of a system, the only processes that can reduce it are gas motions \citep{1990MNRAS.246..678E,1999MNRAS.306..317K,2007ApJ...658..941D}. For local LIRGs, \citet{2008ApJ...674..172R} have proposed that the encounter between two gas rich galaxies causes lower metal abundance gas from the outer regions to fall into the merger central region, providing dilution of the metal abundance. The input of gas powers the star formation and dilutes metals at the same time, explaining their location in the M-Z relationship. 
Independently, \citet{2005A&A...430..115H} have suggested a scenario in which present spiral galaxies are built by the gradual infall of the gas being expelled during the first stages of a major merger. Both scenarios are based on supply of unprocessed gas from the outskirts of encounters during a merger event. For both cases, outflows are unlikely the cause of the gas dilution as they may be dominated by a neutral phase with the metal abundance of the central regions \citep{2008ApJ...674..172R,2005ApJS..160..115R}.

Because distant starbursts \& LIRGs are a common phenomenon in the distant Universe, a merger scenario to explain their properties may have a considerable impact for the formation of present-day spirals. It has prompted by \citet{2005A&A...430..115H} to propose a scenario of disk rebuilding for the majority of spirals galaxies.
Recent simulations of \citet{2008arXiv0805.1246L} indicate that the majority of mergers can reconstruct a low mass star-forming disk. Moreover, the large dispersion in metallicity we observed at intermediate redshifts is qualitatively consistent with the diversity of galaxy building histories within the framework of a merger scenario.  

The external gas supply through merger process is in good agreement with Damped Lyman Absorber (DLA) observations. 
DLAs have a lower metal content compared to emission selected galaxies and have a wide dispersion in metallicity. Absorption measurements through DLAs trace outer regions of galaxies and probably extend envelopes of unprocessed gas on the outskirts of galaxies. \citet{2003ApJ...593..235W} have proposed a scenario for the formation of spiral galaxies from DLAs in which disk galaxies are surrounded by an halo of neutral gas during their assembly. The simulation of \citet{2004ASSL..319..655N} and \citet{2007MNRAS.374.1479G} have reproduced many properties of local spiral galaxies by merger of gas rich encounters.
  
\section{Conclusions}
We have observed 88 intermediate mass galaxies with high quality spectra and
moderate resolution to establish a robust study of the stellar mass
metallicity relation at z$\sim$0.6. Metal abundance have been estimated
following a careful analysis of each individual spectra:  evaluation of
underlying absorption in Balmer lines by stellar continuum subtraction,
robust extinction measurement and evaluation of AGN contamination. 

We confirmed the shift about $\sim$0.3\,dex to lower abundance of the M-Z
relation at z$\sim$0.6 which has been found by \citet{2006A&A...447..113L}. Combined with the robust selection of our sample, the relatively high number of studied galaxies in this study allows to derive the evolution of [12+log(O/H)] over  the
redshift bin $0.4<z<0.9$. It evidences that the median metal abundance of galaxies 
diminishes with the redshift. Using similar measurements from samples at higher redshift we
recover the evolution in metals in the ionised gas from z=0 to z$\sim$3: we find a linear evolution of metallicity as a function of the lookback time. This simply means that the evolution of the gas phase in massive galaxies is still active down to z=0.4 conversely to the popular belief that all massive galaxies have their stellar content locked since z=1.
 
Our results are thus in strong disagreement with those of \citet{2005ApJ...635..260S} et
al. They indeed found that the evolution of galaxies can be at first order explained by a close box model and that galaxies with $\log{M_{\mathrm{stellar}}}>10^{10}\mathrm{M_\odot}$ have already reached local
metallicities at z$\sim$0.6. On contrary we have discarded the close box model as a valid approximation of galaxy evolution. We suggest that this discrepancy is due to the low S/N quality of the \citet{2005ApJ...635..260S} data. We notice however that \citet{2005ApJ...635..260S} results have been widely used to construct metal content evolution simulations, explaining why our results are not reproduced by these models. We are calling for a new generation of modelling able to account for our observations.

We have put into evidence the need for a significant contribution of external gas supply in the metal content evolution of galaxies. Since 6\,Gyrs ago, a significant fraction of massive galaxies have undergone a rapid growth in which 30\% of their present stellar mass has been formed from external gas supply. The addition of gas in galaxies explain the metal deficiency of star-forming galaxies at z$\sim$0.6 at all mass scales. 
The origin of the gas supply is a key question. In the secular scenario, the infalling gas come from pristine gas in the large scale filament structures. The perturbed kinematics and morphology of the intermediate redshift galaxies have lead us to favor the hierarchical scenario where the unprocessed gas could arise from large envelopes of gas lying in the halo of galaxies after merging events. 

We have pointed out that local LIRGs and distant starbursts and LIRGs share similar properties such as star-formation efficiency, metal abundance and morphology. We suggest an unique evolution scenario for both populations, which is linked to the gas infall provided by galaxy merging and interactions. Progenitors of spiral galaxies have their disks rebuilt through an infall of the gas which has been previously expelled in the first stage of a gas rich merger. Local LIRGs are in this case latecomers of the starbursts/LIRGs population predominant at z$\sim$0.6.  A better understanding of the kinematics and metallicity of galactic  halos in gas-rich system at intermediate redshift is needed to disentangle the origin of the external gas supply.

\begin{acknowledgements} This work is supported by a PhD scholarship from Funda\c{c}\~ao para a Ciencia e a Tecnologia, grants from Region Ile-de-France, the National Basic Research Program of China (973 Program) No.2007CB815404 and the Knowledge Innovation Program of the Chinese Academy of Sciences. We want to thank you Ana Mour\~ao for improving the english language in the text. \end{acknowledgements} 
\bibliographystyle{aa} 
 \bibliography{IMAGE-IV} 
 \clearpage \onecolumn
 
 \begin{table} 
\caption{Identified systematic uncertainties that could impact on the shift of the M-Z relation between z$\sim$0.6 and high-z galaxies. } 
\label{budget2} 
\centering
\begin{tabular}{l c c} 
\hline
Possible bias & $ Z$& Comments \\ 
\hline 
Metallicity calibration &0.3& Systhematic between $N2$ and $R_{23}$ calibrations. \\
Properties of \ion{H}{ii} region &-0.16 & For high-z sample with $N2$ calibrations. \\
Extinction &-0.3& $R_{23}$ estimated sample without extinction correction\\
Stellar mass&-0.05& See Sect 5.4 \\
\hline 
\end{tabular} 
\end{table}

 \begin{longtable}{c c c c c c c c r r} 
\caption{\label{table1} Basic data from galaxies in sample~A.  The IR luminosity of galaxies has been estimated from the mid-IR catalogue of \citet{2005ApJ...632..169L} and using the procedure of \citet{2001ApJ...556..562C}. The rest frame magnitude in J,K-band and the luminosity at 2800$\AA$ have been derived by modeling galaxy SEDs using ISAAC and ACS multi-band photometry. }\\ 

\hline
Name  & z & RA (J2000) &  DEC (J2000) & $I_{\mathrm{AB}}$ & $L_{\mathrm{IR}}$ & $M_\mathrm{J}$& $M_\mathrm{K}$ \\[1mm]

\hline 
\endhead 
\hline 
\endfoot 
J033210.92-274722.8 & $0.417$ & $53.04549408$ & $-27.78966522$ & $20.27$ & $10.95$ & $-21.97$ & $-21.89$\\
J033211.70-274507.6 & $0.677$ & $53.04874039$ & $-27.75211334$ & $22.63$ & $-$ & $-19.89$ & $-19.74$\\
J033212.30-274513.1 & $0.645$ & $53.05125809$ & $-27.75362968$ & $21.60$ & $10.90$ & $-21.89$ & $-21.84$\\
J033212.39-274353.6 & $0.422$ & $53.05161285$ & $-27.73155403$ & $21.40$ & $10.94$ & $-21.59$ & $-21.57$\\
J033212.51-274454.8 & $0.732$ & $53.05210876$ & $-27.74856567$ & $22.81$ & $-$ & $-19.87$ & $-19.56$\\
J033213.76-274616.6 & $0.679$ & $53.05731201$ & $-27.77127075$ & $22.92$ & $-$ & $-19.41$ & $-19.07$\\
J033214.48-274320.1 & $0.546$ & $53.06032181$ & $-27.72223854$ & $22.84$ & $-$ & $-19.28$ & $-19.17$\\
J033215.36-274506.9 & $0.860$ & $53.06398392$ & $-27.75192451$ & $22.29$ & $-$ & $-21.57$ & $-21.47$\\
J033217.36-274307.3 & $0.647$ & $53.07234955$ & $-27.71868706$ & $21.30$ & $-$ & $-21.77$ & $-21.67$\\
J033217.75-274547.7 & $0.734$ & $53.07396698$ & $-27.76324844$ & $22.14$ & $-$ & $-21.20$ & $-21.11$\\
J033219.32-274514.0 & $0.725$ & $53.0804863$ & $-27.75389862$ & $22.26$ & $-$ & $-21.25$ & $-21.16$\\
J033219.96-274449.8 & $0.784$ & $53.08317947$ & $-27.74717331$ & $22.60$ & $-$ & $-20.88$ & $-20.78$\\
J033222.13-274344.5 & $0.541$ & $53.09218979$ & $-27.72902679$ & $23.03$ & $-$ & $-19.03$ & $-18.82$\\
J033223.06-274226.3 & $0.734$ & $53.09606552$ & $-27.70730209$ & $22.17$ & $-$ & $-21.43$ & $-21.33$\\
J033223.40-274316.6 & $0.616$ & $53.09751511$ & $-27.7212677$ & $20.84$ & $11.35$ & $-23.00$ & $-22.98$\\
J033224.60-274428.1 & $0.538$ & $53.10250473$ & $-27.74114418$ & $22.05$ & $-$ & $-20.47$ & $-20.38$\\
J033225.26-274524.0 & $0.666$ & $53.10525131$ & $-27.75665855$ & $21.40$ & $-$ & $-21.63$ & $-21.54$\\
J033225.46-275154.6 & $0.672$ & $53.10609818$ & $-27.86517906$ & $21.09$ & $11.27$ & $-22.68$ & $-22.60$\\
J033225.77-274459.3 & $0.833$ & $53.10738373$ & $-27.74981499$ & $22.86$ & $11.16$ & $-20.94$ & $-20.71$\\
J033226.21-274426.3 & $0.495$ & $53.10919571$ & $-27.74064445$ & $23.28$ & $-$ & $-18.59$ & $-18.38$\\
J033227.36-275015.9 & $0.769$ & $53.11399078$ & $-27.83775711$ & $21.83$ & $11.01$ & $-21.98$ & $-21.89$\\
J033227.93-274353.6 & $0.458$ & $53.11636734$ & $-27.73156548$ & $23.53$ & $-$ & $-17.86$ & $-17.59$\\
J033227.93-275235.6 & $0.383$ & $53.11637878$ & $-27.87656403$ & $20.55$ & $10.69$ & $-21.59$ & $-21.50$\\
J033229.32-275155.4 & $0.510$ & $53.12216568$ & $-27.86540031$ & $22.25$ & $-$ & $-19.42$ & $-19.09$\\
J033229.64-274242.6 & $0.667$ & $53.12350845$ & $-27.71182251$ & $20.90$ & $11.77$ & $-22.74$ & $-22.66$\\
J033229.71-274507.2 & $0.737$ & $53.12377548$ & $-27.75200462$ & $22.39$ & $-$ & $-20.93$ & $-20.82$\\
J033230.07-274534.2 & $0.648$ & $53.12527084$ & $-27.75950241$ & $21.71$ & $-$ & $-21.44$ & $-21.34$\\
J033230.57-274518.2 & $0.679$ & $53.12736893$ & $-27.75506783$ & $20.81$ & $11.41$ & $-23.00$ & $-22.91$\\
J033231.58-274612.7 & $0.654$ & $53.13157272$ & $-27.77020454$ & $22.44$ & $-$ & $-20.35$ & $-20.26$\\
J033232.13-275105.5 & $0.682$ & $53.13389587$ & $-27.85154152$ & $22.34$ & $11.07$ & $-21.71$ & $-21.68$\\
J033232.32-274343.6 & $0.534$ & $53.1346817$ & $-27.72879028$ & $22.84$ & $-$ & $-19.26$ & $-19.15$\\
J033232.58-275053.9 & $0.670$ & $53.13574982$ & $-27.84830666$ & $21.86$ & $-$ & $-21.54$ & $-21.46$\\
J033233.00-275030.2 & $0.669$ & $53.13750839$ & $-27.84171295$ & $21.43$ & $11.62$ & $-23.18$ & $-23.15$\\
J033233.82-274410.0 & $0.666$ & $53.1409111$ & $-27.73611832$ & $21.51$ & $-$ & $-22.61$ & $-22.55$\\
J033233.90-274237.9 & $0.619$ & $53.14123917$ & $-27.71053696$ & $21.22$ & $10.92$ & $-21.84$ & $-21.74$\\
J033234.04-275009.7 & $0.703$ & $53.14182281$ & $-27.83602524$ & $22.35$ & $-$ & $-20.69$ & $-20.60$\\
J033234.88-274440.6 & $0.677$ & $53.14534378$ & $-27.74459839$ & $23.13$ & $11.21$ & $-20.65$ & $-20.56$\\
J033234.91-274501.9 & $0.665$ & $53.14544678$ & $-27.75053024$ & $22.48$ & $-$ & $-20.42$ & $-20.31$\\
J033236.37-274543.3 & $0.435$ & $53.15153885$ & $-27.76201439$ & $22.14$ & $10.88$ & $-21.04$ & $-21.01$\\
J033236.52-275006.4 & $0.689$ & $53.15215683$ & $-27.83511353$ & $21.65$ & $10.98$ & $-21.78$ & $-21.69$\\
J033236.72-274406.4 & $0.666$ & $53.1529808$ & $-27.73512459$ & $22.07$ & $11.05$ & $-22.02$ & $-21.94$\\
J033236.74-275206.9 & $0.784$ & $53.15309906$ & $-27.86857033$ & $22.30$ & $11.12$ & $-21.39$ & $-21.30$\\
J033237.26-274610.3 & $0.736$ & $53.15524292$ & $-27.76953125$ & $22.38$ & $-$ & $-21.15$ & $-21.06$\\
J033237.49-275216.1 & $0.423$ & $53.15618896$ & $-27.87112999$ & $21.09$ & $10.47$ & $-20.87$ & $-20.78$\\
J033237.96-274652.0 & $0.620$ & $53.15816498$ & $-27.78109741$ & $21.97$ & $-$ & $-20.53$ & $-20.43$\\
J033238.77-274732.1 & $0.458$ & $53.16156006$ & $-27.79225922$ & $21.13$ & $11.48$ & $-21.32$ & $-21.24$\\
J033238.97-274630.2 & $0.420$ & $53.16236496$ & $-27.77506256$ & $21.07$ & $10.35$ & $-21.22$ & $-21.13$\\
J033240.04-274418.6 & $0.523$ & $53.16683197$ & $-27.73850822$ & $20.89$ & $10.89$ & $-22.02$ & $-21.93$\\
J033240.32-274722.8 & $0.619$ & $53.16801453$ & $-27.78967285$ & $23.00$ & $-$ & $-19.62$ & $-19.53$\\
J033243.96-274503.5 & $0.533$ & $53.1831665$ & $-27.75096512$ & $22.41$ & $-$ & $-20.16$ & $-20.07$\\
J033244.44-274819.0 & $0.416$ & $53.18515015$ & $-27.80527496$ & $20.57$ & $10.79$ & $-22.02$ & $-21.95$\\
J033245.11-274724.0 & $0.436$ & $53.18795013$ & $-27.78999901$ & $20.81$ & $-$ & $-22.08$ & $-22.05$\\
J033245.51-275031.0 & $0.562$ & $53.18963242$ & $-27.84193993$ & $22.70$ & $-$ & $-19.52$ & $-19.43$\\
J033245.63-275133.0 & $0.858$ & $53.19010544$ & $-27.85918045$ & $22.40$ & $-$ & $-21.02$ & $-20.68$\\
J033245.78-274812.9 & $0.534$ & $53.19076157$ & $-27.80357933$ & $21.70$ & $-$ & $-21.56$ & $-21.51$\\
J033248.84-274531.5 & $0.278$ & $53.20350266$ & $-27.75874901$ & $21.50$ & $9.81$ & $-19.56$ & $-19.47$\\
J033249.58-275203.1 & $0.415$ & $53.20658112$ & $-27.86752319$ & $21.14$ & $10.70$ & $-21.48$ & $-21.45$\\
J033252.85-275207.9 & $0.684$ & $53.22020721$ & $-27.86885071$ & $23.06$ & $11.10$ & $-20.64$ & $-20.57$\\
 \end{longtable} 
 
 \begin{longtable}{c c c c c c c } 
\caption{\label{table2} The extinction $A_{\mathrm{V}}$(Balmer), $A_{\mathrm{V}}$(IR) and the adopted $A_{\mathrm{V}}$. The $A_{\mathrm{V}}$(IR) calculated with the lower limit of IR detection are upper limit. The final extinction estimated only with the limit $L_{\mathrm{IR}}$ are noted with $^1$. Three SFR are given : $SFR_{\mathrm{IR}}$, $SFR_{\mathrm{Hb}}$, $SFR_{\mathrm{UV}}$. The $SFR_{\mathrm{Hb}}$ is corrected from extinction and aperture. To determine the unobscured $SFR_{\mathrm{UV}}$, we used the rest-frame 2800$\AA$ luminosity from SED fitting }\\ 
\hline 
Name  & $A_{\mathrm{V}}$& $A_{\mathrm{V}}$& $A_{\mathrm{V}}$&$SFR_{\mathrm{IR}}$ &$SFR_{\mathrm{Hb}}$&$SFR_{\mathrm{UV}}$  \\[1mm]
&Balmer &IR  &  &$M_\odot\,yr^-{1}$ & $M_\odot\,yr^-{1}$&$M_\odot\,yr^-{1}$\\[1mm]
\hline 
\endhead 
\hline 
\endfoot 
J033210.92-274722.8&$0.77^{+0.17}_{-0.17}$&$1.27\pm0.32$&$1.02^{+0.24}_{-0.25}$&$15.14\pm7.50$&$11.08^{+3.82}_{-4.00}$&4.60\\[1mm]
J033211.70-274507.6&-&$<0.31$&$0.31^1$&$<16.51$&$<16.58$&2.04\\[1mm]
J033212.30-274513.1&$1.85^{+0.39}_{-0.41}$&$1.29\pm0.31$ &$1.57^{+0.35}_{-0.36}$&$13.43\pm6.66$&$18.91^{+10.22}_{-10.58}$&3.37\\[1mm]
J033212.39-274353.6&$0.08^{+1.13}_{-0.08}$&$2.44\pm0.30$ &$0.08^{+1.13}_{-0.08}$&$14.95\pm4.51$&$3.49^{+4.89}_{-3.49}$&1.03\\[1mm]
J033212.51-274454.8&$1.79^{+0.40}_{-0.42}$&$<0.05$&$0.87^{+0.35}_{-0.37}$&$<20.66$&$64.07^{+34.60}_{-37.07}$&2.90\\[1mm]
J033213.76-274616.6&-&$<1.97$&$1.97^1$&$<16.66$&$<16.65$&1.45\\[1mm]
J033214.48-274320.1&$1.35^{+1.15}_{-1.35}$ &$<2.603$&$1.98^{+0.72}_{-0.84}$ &$<9.05$&$4.19^{+6.00}_{-4.19}$&0.87\\[1mm]
J033215.36-274506.9&$2.02^{+0.20}_{-0.20}$&$<1.077$&$1.55\pm0.26$&$<32.99$&$59.06^{+22.34}_{-22.34}$&4.38\\[1mm]
J033217.36-274307.3&-&$<0.28$&$0.28^1$&$<14.56$&$<14.59$&5.83\\[1mm]
J033217.75-274547.7&-&$<1.70$&$1.70^1$&$<20.87$&$<20.79$&3.21\\[1mm]
J033219.32-274514.0&$2.86^{+0.65}_{-0.72}$&$<1.46$&$2.16^{+0.47}_{-0.50}$&$<20.11$&$47.95^{+37.68}_{-40.91}$&2.81\\[1mm]
J033219.96-274449.8&-&$<2.17$&$2.17^1$&$<25.17$&$<25.25$&2.30\\[1mm]
J033222.13-274344.5&$1.53^{+1.29}_{-1.53}$&$<3.22$&$2.38^{+0.78}_{-0.92}$&$<8.83$&$3.13^{+5.06}_{-3.13}$&0.56\\[1mm]
J033223.06-274226.3&-&$<1.91$ &$1.91^1$&$<20.84$&$<20.77$&2.92\\[1mm]
J033223.40-274316.6&$0.00^{+0.57}$&$2.04\pm0.30$&$1.02^{+0.44}_{-0.64}$&$37.95\pm18.80$&$10.79^{+7.78}_{-10.79}$&4.25\\[1mm]
J033224.60-274428.1&$2.79^{+0.76}_{-0.85}$&$<1.86$&$2.33^{+0.53}_{-0.57}$&$<8.70$&$15.44^{+14.25}_{-15.44}$&1.65\\[1mm]
J033225.26-274524.0&-&$<1.49$&$1.49^1$&$<15.79$&$<15.84$&5.07\\[1mm]
J033225.46-275154.6&-&$1.71\pm0.29$&$1.71\pm0.29$&$32.10\pm15.90$&$31.96^{+13.75}_{-13.75}$&5.92\\[1mm]
J033225.77-274459.3&-&$0.71\pm0.32$&$0.71\pm0.32$&$24.86\pm12.32$&$24.92^{+12.07}_{-12.07}$&2.96\\[1mm]
J033226.21-274426.3&$2.96^{+1.98}_{-2.77}$&$<2.48$&$2.72^{+1.13}_{-1.52}$&$<6.92$&$9.36^{+28.38}_{-9.36}$&0.51\\[1mm]
J033227.36-275015.9&-&$0.79\pm0.31$&$0.79\pm0.31$&$17.49\pm8.67$&$17.49\pm8.15$&4.63\\[1mm]
J033227.93-274353.6&$0.00^{+0.27}$&$<2.02$&$1.01^{+0.28}_{-1.06}$&$<5.61$&$1.62^{+0.67}_{-1.62}$&0.32\\[1mm]
J033227.93-275235.6&$0.32^{+0.12}_{-0.12}$&$1.16\pm0.32$&$0.74\pm0.22$&$8.46\pm4.19$&$5.05^{+1.58}_{-1.58}$&4.31\\[1mm]
J033229.32-275155.4&$1.31^{+0.58}_{-0.64}$&$<1.07$&$1.19^{+0.44}_{-0.47}$&$<7.52$&$8.68^{+6.26}_{-6.82}$&2.18\\[1mm]
J033229.64-274242.6&-&$1.96\pm0.32$&$1.96\pm0.32$&$101.08\pm50.08$&$100.71^{+48.76}_{-48.76}$&7.04\\[1mm]
J033229.71-274507.2&$2.05^{+0.63}_{-0.69}$&$<1.61$&$1.83^{+0.47}_{-0.50}$&$<21.08$&$27.74^{+21.80}_{-23.67}$&2.75\\[1mm]
J033230.07-274534.2&$1.24^{+0.23}_{-0.23}$&$<1.17$&$1.21\pm0.27$&$<14.59$&$15.27^{+6.04}_{-6.04}$&3.24\\[1mm]
J033230.57-274518.2&$1.24^{+0.24}_{-0.25}$&$1.67\pm0.31$&$1.45\pm0.28$&$43.69\pm21.65$&$33.30^{+13.74}_{-13.74}$&11.19\\[1mm]
J033231.58-274612.7&$3.10^{+0.29}_{-0.31}$&$<1.39$&$2.25\pm0.31$&$<15.00$&$43.41^{+20.22}_{-20.22}$&1.96\\[1mm]
J033232.13-275105.5&$0.18^{+1.43}_{-0.18}$&$2.55\pm0.31$&$0.18^{+1.43}_{-0.18}$&$20.13\pm9.98$&$4.64^{+8.93}_{-4.64}$&1.40\\[1mm]
J033232.32-274343.6&-&$<1.73$&$1.73^1$&$<8.52$&$<8.48$&0.60\\[1mm]
J033232.58-275053.9&-&$<1.53$&$1.53^1$&$<16.05$&$<16.00$&3.15\\[1mm]
J033233.00-275030.2&-&$2.41\pm0.32$&$2.41\pm0.32$&$71.34\pm35.35$&$70.93^{+34.34}_{-34.34}$&2.38\\[1mm]
J033233.82-274410.0&-&$<2.19$&$2.19^1$&$<15.79$&$<15.80$&1.42\\[1mm]
J033233.90-274237.9&$0.62^{+0.25}_{-0.26}$&$1.13\pm0.32$&$0.88^{+0.28}_{-0.29}$&$14.25\pm7.06$&$10.47^{+4.32}_{-4.51}$&5.71\\[1mm]
J033234.04-275009.7&$0.65^{+0.50}_{-0.54}$&$<1.64$&$1.14^{+0.40}_{-0.42} $&$<18.39$&$9.93^{+6.34}_{-6.75}$&2.87\\[1mm]
J033234.88-274440.6&-&$2.95\pm0.28$&$2.95\pm0.28$&$27.90\pm13.82$&$28.00^{+11.55}_{-11.55}$&0.34\\[1mm]
J033234.91-274501.9&-&$<2.01$&$2.01^1$&$<15.74$&$<15.76$&1.88\\[1mm]
J033236.37-274543.3&-&$5.27\pm0.28$&$5.27\pm0.28$&$12.76\pm6.32$&$12.69\pm5.23$&0.62\\[1mm]
J033236.52-275006.4&$1.70^{+0.42}_{-0.45}$&$1.25\pm0.32$&$1.48^{+0.37}_{-0.38}$&$16.50\pm8.17$&$21.88^{+12.66}_{-13.09}$&4.78\\[1mm]
J033236.72-274406.4&$2.90^{+1.33}_{-1.65}$&$2.08\pm0.23$&$2.49^{+0.78}_{-0.94}$&$19.17\pm9.50$&$31.69^{+51.27}_{-31.69}$&1.17\\[1mm]
J033236.74-275206.9&-&$1.66\pm0.31$&$1.66\pm0.31$&$22.68\pm11.24$&$22.57^{+10.51}_{-10.51}$&3.05\\[1mm]
J033237.26-274610.3&$3.25^{+0.53}_{-0.57}$&$<1.74$&$2.49^{+0.42}_{-0.44}$&$<21.02$&$53.23^{+36.15}_{-38.38}$&2.44\\[1mm]
J033237.49-275216.1&$1.36^{+0.20}_{-0.20}$&$0.78\pm0.32$&$1.07\pm0.26$&$5.04\pm2.50$&$7.16^{+2.71}_{-2.71}$&3.19\\[1mm]
J033237.96-274652.0&$1.80^{+0.64}_{-0.70}$&$<1.28$&$1.54^{+0.47}_{-0.50}$&$<12.89$&$17.67^{+13.89}_{-15.08}$&3.44\\[1mm]
J033238.77-274732.1&$2.36^{+0.15}_{-0.15}$&$1.68\pm0.32$&$2.02\pm0.24$&$51.86\pm25.70$&$78.58^{+27.08}_{-27.08}$&5.33\\[1mm]
J033238.97-274630.2&$0.00^{+1.13}$&$0.92\pm0.31$&$0.46^{+0.72}_{-0.46}$&$3.83\pm1.90$&$2.18^{+3.12}_{-1.67}$&2.18\\[1mm]
J033240.04-274418.6&-&$2.21\pm0.26$&$2.21\pm0.26$&$13.26\pm6.57$&$13.28^{+5.02}_{-5.02}$&1.68\\[1mm]
J033240.32-274722.8&$2.29^{+1.10}_{-1.30}$&$<2.18$&$2.24^{+0.68}_{-0.79}$&$<12.85$&$13.81^{+18.14}_{-13.81}$&1.06\\[1mm]
J033243.96-274503.5&$4.51^{+0.93}_{-1.07}$&$<1.79$&$3.15^{+0.62}_{-0.69}$&$<8.47$&$45.28^{+52.02}_{-45.28}$&1.15\\[1mm]
J033244.44-274819.0&$1.20^{+0.26}_{-0.27}$&$1.24\pm0.32$&$1.22\pm0.29$&$10.62\pm5.26$&$10.41^{+4.48}_{-4.48}$&2.46\\[1mm]
J033245.11-274724.0&$2.10^{+0.39}_{-0.41}$&$<0.84$&$1.15^{+0.34}_{-0.36}$&$20.8$&$7.16^{+3.73}_{-4.01}$&1.93\\[1mm]
J033245.51-275031.0&$2.67^{+0.77}_{-0.87}$&$<2.84$&$2.75^{+0.54}_{-0.58}$&$<9.80$&$8.73^{+8.27}_{-8.73}$&0.78\\[1mm]
J033245.63-275133.0&$0.35^{+0.34}_{-0.35}$&$<1.79$&$1.07^{+0.33}_{-0.33}$&$<32.70$&$13.53^{+6.80}_{-6.80}$&4.13\\[1mm]
J033245.78-274812.9&-&$<2.18$&$2.18^1$&$<8.52$&$<8.54$&0.72\\[1mm]
J033248.84-274531.5&$0.96^{+0.62}_{-0.68}$&$1.02\pm0.31$&$0.99^{+0.46}_{-0.49}$&$1.11\pm0.55$&$1.06^{+0.81}_{-0.88}$&0.36\\[1mm]
J033249.58-275203.1&$2.03^{+0.48}_{-0.52}$&$1.78\pm0.32$&$1.90^{+0.40}_{-0.42} $&$8.56\pm4.24$&$9.93^{+6.34}_{-6.74}$&1.32\\[1mm]
J033252.85-275207.9&-&$2.20\pm0.31$&$2.20\pm0.31 $&$21.64\pm10.72$&$21.55^{+10.04}_{-10.04}$&0.64\\[1mm]

 \end{longtable}

\begin{longtable}{c c c c  c c r r} 
\caption{\label{table3} Some important emission line ratios, the oxygen abundance in ISM and stellar mass. The uncertainties of the line ratio and metallicity are from uncertainties of extinction and emission line flux measurement. Galaxies in which extinction have been estimated only with $A_V(IR_{\mathrm{lim}}$) have lower limit of metallicity.  When log$R_{23}>1$ the oxygen abundance is quoted with $^1$)}\\ 
\hline
& &  &  &   & & & \\
Name  & z &  $log{\frac{[\ion{O}{ii}]}{\mathrm{H\beta}}}$ & $log{\frac{[\ion{O}{iii}]}{\mathrm{H\beta}}}$ & $log{\frac{[\ion{Ne}{iii}]}{[\ion{O}{ii}]}}$ & $12+log{O/H}$ & $log{\frac{M}{\mathrm{M_\odot}}}$\\
& &  &  &   & & & \\

\hline 
\endhead 
\hline 
\endfoot 

J033210.92-274722.8&0.416&$0.36\pm0.02$&$-0.14\pm0.01$&-&$8.94\pm0.01$&$10.68$\\[1mm]
J033211.70-274507.6&0.676&$0.19\pm0.01$&$0.65\pm0.01$&-&$>8.63$&$9.76$\\[1mm]
J033212.30-274513.1&0.645&$0.23\pm0.04$&$-0.23\pm0.04$&-&$9.02\pm0.03$&$10.63$\\[1mm]
J033212.39-274353.6&0.422&$0.43\pm0.52$&$0.07\pm0.06$&-&$8.83\pm0.35$&$10.58$\\[1mm]
J033212.51-274454.8&0.732&$0.14\pm0.04$&$0.75\pm0.01$&-0.66&$8.54\pm0.01$&$9.67$\\[1mm]
J033213.76-274616.6&0.679&$0.57\pm0.02$&$0.5\pm0.01$&-&$>8.55$&$9.50$\\[1mm]
J033214.48-274320.1&0.546&$0.94\pm0.21$&$0.51\pm0.03$&-&$8.29^1$&$9.56$\\[1mm]
J033215.36-274506.9&0.860&$0.54\pm0.25$&$0.18\pm0.01$&-1.82&$8.73\pm0.19$&$10.46$\\[1mm]
J033217.36-274307.3&0.647&$0.27\pm0.19$&$0.17\pm0.01$&-&$>8.90$&$10.54$\\[1mm]
J033217.75-274547.7&0.734&$0.58\pm0.04$&$0.22\pm0.01$&-&$>8.68$&$10.31$\\[1mm]
J033219.32-274514.0&0.725&$0.95\pm0.04$&$0.26\pm0.02$&-1.56&$8.29^1$&$10.33$\\[1mm]
J033219.96-274449.8&0.784&$0.62\pm0.04$&$0.01\pm0.03$&-&$>8.70$&$10.19$\\[1mm]
J033222.13-274344.5&0.541&$1.05\pm0.17$&$0.6\pm0.02$&-&$8.29^1$&$9.43$\\[1mm]
J033223.06-274226.3&0.734&$0.81\pm0.14$&$0.06\pm0.12$&-&$>8.48$&$10.41$\\[1mm]
J033223.40-274316.6&0.615&$0.26\pm0.06$&$-0.33\pm0.03$&-&$9.02\pm0.03$&$11.11$\\[1mm]
J033224.60-274428.1&0.538&$0.88\pm0.1$&$0.43\pm0.02$&-&$8.29^1$&$10.05$\\[1mm]
J033225.26-274524.0&0.666&$0.49\pm0.01$&$0.21\pm0.02$&-1.22&$>8.76$&$10.50$\\[1mm]
J033225.46-275154.6&0.672&$0.37\pm0.04$&$-0.05\pm0.07$&-1.47&$8.91\pm0.04$&$10.94$\\[1mm]
J033225.77-274459.3&0.832&$0.02\pm0.05$&$0.61\pm0.01$&-&$8.72\pm0.01$&$10.14$\\[1mm]
J033226.21-274426.3&0.495&-&$0.65\pm0.06$&-&$>8.31$&$9.24$\\[1mm]
J033227.36-275015.9&0.769&$0.14\pm0.01$&$-0.2\pm0.02$&-1.31&$9.06\pm0.01$&$10.63$\\[1mm]
J033227.93-274353.6&0.458&$0.44\pm0.04$&$0.35\pm0.01$&-&$8.73\pm0.02$&$8.95$\\[1mm]
J033227.93-275235.6&0.383&$-0.01\pm0.03$&$-0.52\pm0.01$&-&$9.15\pm0.01$&$10.52$\\[1mm]
J033229.32-275155.4&0.510&$0.85\pm0.07$&$0.58\pm0.02$&-1.23&$8.29^1$&$9.52$\\[1mm]
J033229.64-274242.6&0.667&$0.5\pm0.05$&$-0.19\pm0.04$&-&$8.85\pm0.04$&$10.96$\\[1mm]
J033229.71-274507.2&0.737&$0.44\pm0.09$&$-0.08\pm0.02$&-&$8.87\pm0.06$&$10.19$\\[1mm]
J033230.07-274534.2&0.648&$0.54\pm0.07$&$0.22\pm0.01$&-&$8.71\pm0.06$&$10.43$\\[1mm]
J033230.57-274518.2&0.679&$0.1\pm0.03$&$-0.31\pm0.01$&-&$9.09\pm0.01$&$11.05$\\[1mm]
J033231.58-274612.7&0.654&$0.57\pm0.04$&$0.47\pm0.01$&-1.03&$8.57\pm0.03$&$9.97$\\[1mm]
J033232.13-275105.5&0.682&$0.39\pm0.83$&$0.02\pm0.04$&-1.21&$8.84\pm0.59$&$10.58$\\[1mm]
J033232.32-274343.6&0.534&$0.41\pm0.1$&$0.26\pm0.13$&-&$>8.79$&$9.56$\\[1mm]
J033232.58-275053.9&0.670&$0.36\pm0.01$&$-0.32\pm0.01$&-&$>8.97$&$10.46$\\[1mm]
J033233.00-275030.2&0.669&$0.28\pm0.04$&$0.29\pm0.02$&-1.19&$>8.85$&$11.19$\\[1mm]
J033233.82-274410.0&0.666&$0.51\pm0.01$&$0.48\pm0.1$&-&$>8.61$&$10.97$\\[1mm]
J033233.90-274237.9&0.619&$0.47\pm0.03$&$0.01\pm0.01$&-&$8.83\pm0.02$&$10.58$\\[1mm]
J033234.04-275009.7&0.703&$0.55\pm0.05$&$0.37\pm0.02$&-&$8.64\pm0.03$&$10.10$\\[1mm]
J033234.88-274440.6&0.677&$0.64\pm0.04$&$0.27\pm0.04$&-&$8.60\pm0.04$&$10.16$\\[1mm]
J033234.91-274501.9&0.665&$0.61\pm0.01$&$-0.02\pm0.05$&-&$>8.72$&$10.00$\\[1mm]
J033236.37-274543.3&0.435&-&$0.13\pm0.01$&-&$8.76\pm0.01$&$10.35$\\[1mm]
J033236.52-275006.4&0.689&$0.3\pm0.05$&$-0.04\pm0.01$&-&$8.95\pm0.03$&$10.55$\\[1mm]
J033236.72-274406.4&0.666&$0.84\pm0.14$&$0.65\pm0.1$&-1.14&$8.29^1$&$10.71$\\[1mm]
J033236.74-275206.9&0.784&$0.64\pm0.03$&$0.09\pm0.01$&-&$8.67\pm0.03$&$10.39$\\[1mm]
J033237.26-274610.3&0.736&$0.38\pm0.07$&$0.11\pm0.06$&-&$8.86\pm0.06$&$10.29$\\[1mm]
J033237.49-275216.1&0.423&$0.48\pm0.02$&$0.46\pm0.04$&-1.57&$8.64\pm0.04$&$10.22$\\[1mm]
J033237.96-274652.0&0.619&$0.63\pm0.07$&$0.47\pm0.02$&-&$8.51\pm0.05$&$10.04$\\[1mm]
J033238.77-274732.1&0.458&-&$0.06\pm0.01$&-&$8.72\pm0.01$&$10.39$\\[1mm]
J033238.97-274630.2&0.419&-&$-0.34\pm0.01$&-&$9.02\pm0.01$&$10.38$\\[1mm]
J033240.04-274418.6&0.523&$0.91\pm0.1$&$0.45\pm0.14$&-1.04&$8.29^1$&$10.72$\\[1mm]
J033240.32-274722.8&0.619&$0.75\pm0.14$&$0.25\pm0.05$&-&$8.50\pm0.14$&$9.69$\\[1mm]
J033243.96-274503.5&0.533&$0.9\pm0.16$&$1.01\pm0.02$&-&$8.29^1$&$9.93$\\[1mm]
J033244.44-274819.0&0.416&-&$-0.26\pm0.01$&-&$8.91\pm0.01$&$10.73$\\[1mm]
J033245.11-274724.0&0.436&$-0.01\pm0.01$&$-0.47\pm0.07$&-&$9.14\pm0.01$&$10.77$\\[1mm]
J033245.51-275031.0&0.562&$0.81\pm0.11$&$0.77\pm0.01$&-0.94&$8.29^1$&$9.68$\\[1mm]
J033245.63-275133.0&0.858&$0.56\pm0.04$&$0.24\pm0.04$&-1.21&$8.69\pm0.05$&$10.15$\\[1mm]
J033245.78-274812.9&0.534&$0.21\pm0.02$&$-0.1\pm0.01$&-&$>9.01$&$10.56$\\[1mm]
J033248.84-274531.5&0.278&$0.36\pm0.08$&$0.38\pm0.01$&-1.18&$8.76\pm0.04$&$9.74$\\[1mm]
J033249.58-275203.1&0.415&$0.36\pm0.06$&$-0.31\pm0.01$&-&$8.97\pm0.04$&$10.53$\\[1mm]
J033252.85-275207.9&0.684&$0.76\pm0.03$&$0.66\pm0.01$&-0.89&$8.29^1$&$10.14$\\[1mm]

\end{longtable}

 \end{document}